\documentclass[useAMS,usenatbib]{mn2e}

\usepackage{psfig,epsfig}
\usepackage{mathrsfs}
\usepackage{amssymb}
\usepackage{epsfig}
\usepackage{marvosym}
\usepackage{subfigure}
\usepackage{fancyhdr}
\usepackage{rotating}

\def\degr{\hbox{$^\circ$}}

\def\arcsec{\hbox{$^{\prime\prime}$}}

\def\farcs{\hbox{$.\!\!^{\prime\prime}$}}
\def\gsim{\mathrel{\hbox{\rlap{\lower.55ex \hbox {$\sim$}}
                   \kern-.3em \raise.4ex \hbox{$>$}}}}
\def\lsim{\mathrel{\hbox{\rlap{\lower.55ex \hbox {$\sim$}}
                   \kern-.3em \raise.4ex \hbox{$<$}}}}

\def\he2{\hbox{He\,{\sc ii} $\lambda$4686}}

\def\UVEX{\hbox{\sl UVEX\ }}
\def\IPHAS{\hbox{\sl IPHAS\ }}
\def\RL1{\hbox{{$R_{L_{1}}$}}}

\def\heir{\hbox{$He${\sc i}$-$$r$\,}}

\title[The UV-Excess Survey of the Northern Galactic Plane (UVEX)]{The UV-Excess survey of the
  Northern Galactic Plane \\(UVEX)}
\author[Paul J. Groot et al.]{{Paul J. Groot$^{1}$\thanks{E-mail:
p.groot@astro.ru.nl},
Kars Verbeek$^{1}$,
Robert Greimel$^{2,3}$,
Mike Irwin$^{4}$,}
\newauthor{
Eduardo Gonz{\'a}lez-Solares$^{4}$,
Boris T. G\"ansicke$^{5}$,
Eelco de Groot$^{1}$, 
Janet Drew$^{6}$,}
\newauthor{
Thomas Augusteijn$^{7}$,
Amornrat Aungwerojwit$^{5,8}$,
Mike Barlow$^{9}$,
Susana Barros$^{5}$,
}
\newauthor{
Else J.M. van den Besselaar$^{1}$,
Jorge Casares$^{10}$,
Romano Corradi$^{2,10}$, 
}
\newauthor{
Jes{\'u}s M. Corral-Santana$^{10}$,
Niall Deacon$^{1}$, 
Wilbert van Ham$^{1}$, 
Haili Hu$^{1}$,}
\newauthor{
Uli Heber$^{11}$,
Peter G. Jonker$^{12,13}$,
Rob King$^{14}$,
Christian Knigge$^{15}$,
}
\newauthor{
Antonio Mampaso$^{10}$,
Tom R. Marsh$^{5}$,
Luisa Morales-Rueda$^{1}$,
Ralf Napiwotzki$^{6}$,}
\newauthor{
Tim Naylor$^{14}$,
Gijs Nelemans$^{1}$,
Tim Oosting$^{1}$,
Stylianos Pyrzas$^{2,3}$, 
}
\newauthor{
Magaretha Pretorius$^{16}$,
Pablo Rodr{\'\i}guez-Gil$^{2,10}$,
Gijs H.A. Roelofs$^{13}$, 
Stuart Sale$^{17}$, 
}
\newauthor{
Pim Schellart$^{1}$,
Danny Steeghs$^{5,13}$,
Cezary Szyszka$^{18}$,
Yvonne Unruh$^{17}$, 
}
\newauthor{ 
Nicholas A. Walton$^{4}$,
Simon Weston$^{6}$,
Andrew Witham$^{15}$,
Patrick Woudt$^{19}$ 
}
\newauthor{
and Albert Zijlstra$^{20}$}\\
$^{1}$Department of Astrophysics, IMAPP, Radboud University Nijmegen,
  P.O. Box 9010, 6500 GL Nijmegen, The Netherlands\\
$^{2}$Isaac Newton Group of Telescopes, Apartado de Correos 321,
  E-38700 Santa Cruz de La Palma, Canary Islands, Spain\\
$^{3}$Institut f\"ur Physik, Karl-Franzen Universit\"at Graz,
Universit\"atsplatz 5, 8010 Graz, Austria\\
$^4$Cambridge Astronomy Survey Unit, Institute of Astronomy, University of
  Cambridge, Madingley Road, Cambridge, CB3 0HA, UK\\
$^{5}$Physics Department, University of Warwick, Coventry, CV4 7AL,
  UK\\
$^{6}$Centre for Astronomy Research, Science \& Technology Research
  Institute, University of Hertfordshire, Hatfield, AL10 9AB, UK\\
$^{7}$Nordic Optical Telescope, Apartado 474, E-38700 Santa Cruz de La
  Palma, Canary Islands, Spain\\
$^{8}$Department of Physics, Faculty of Science, Naresuan
University, Phitsanulok, 65000, Thailand\\
$^{9}$Department of Physics and Astronomy, University College London,
Gower Street, London, WC1E 6BT, UK\\
$^{10}$Instituto de Astrof\'{\i}sica de Canarias, C/ Via Lactea, s/n
  E38205- La Laguna (Tenerife), Spain\\
$^{11}$Dr. Remeis-Sternwarte Bamberg, Universit\"at Erlangen-N\"urnberg,
  Sternwartstrasse 7, D-96049 Bamberg, Germany\\
$^{12}$SRON, Netherlands Institute for Space Research, Sorbonnelaan 2, 3584
  CA Utrecht, The Netherlands\\
$^{13}$Harvard-Smithsonian Center for Astrophysics, 60 Garden Street,
  02138 MA, USA\\
$^{14}$School of Physics, University of Exeter, EX4 4QL, UK\\
$^{15}$School of Physics and Astronomy, University of Southampton,
  Southampton, Hampshire, SO17 1BJ, UK\\
$^{16}$South African Astronomical Observatory, Observatory, Cape
Town, South Africa\\
$^{17}$Imperial College of Science, Technology and Medicine, Blackett
Laboratory, London, SW7 2AZ, UK\\
$^{18}$European Southern Observatory, Karl-Schwarzschild-Strasse 2,
D-85748 Garching bei M\"unchen, Germany\\
$^{19}$Department of Astronomy, University of Cape Town, Private Bag,
  Rondebosch 7700, Republic of South Africa\\
$^{20}$ Jodrell Bank Center for Astrophysics, School of Physics and
  Astronomy, University of Manchester, Manchester, UK
}

\begin{document}

\date{Accepted 2009 June 18.  Received 2009 June 17; in original form 2009 May 18}

\pagerange{\pageref{firstpage}--\pageref{lastpage}} \pubyear{2009}

\maketitle

\label{firstpage}

\begin{abstract}
  The UV-Excess Survey of the Northern Galactic Plane images a
  10\degr$\times$185\degr\ wide band, centered on the Galactic Equator
  using the 2.5m Isaac Newton Telescope in four bands ($U,g,r,He${\sc
    i}$5875$) down to $\sim$21$^{st}$-22$^{\rm nd}$ magnitude
  ($\sim$20$^{\rm th}$ in \mbox{$He${\sc i}$5875$}). The setup and data
  reduction procedures are described. Simulations of the colours of
  main-sequence stars, giant, supergiants, DA and DB white dwarfs and
  AM CVn stars are made, including the effects of reddening. A first
  look at the data of the survey (currently 30\% complete) is given.
\end{abstract}


\section{Introduction}

\begin{figure*}
\centerline{\psfig{figure=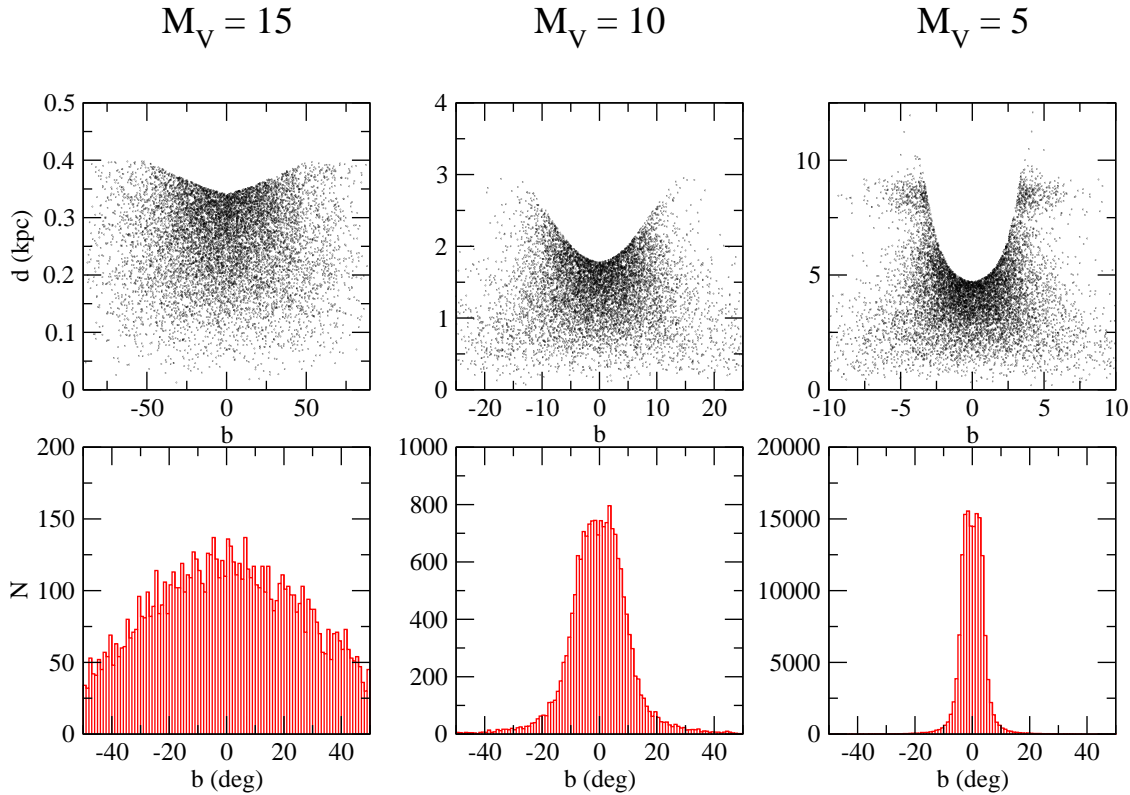,width=12cm,angle=-90}}
\caption{Distribution as a function of heliocentric distance vs.
  latitude (upper panels) and a histrogram of number of systems vs.
  latitude (lower panels) of three Galactic populations with absolute
  visual magnitudes M$_V$ = 15, 10 and 5 (left to right), based on a
  Galaxy model according to Boissier \&Prantzos (1999), and including
  a Sandage (1972) extinction model, as detailed in Nelemans et al.
  (2004). A sample with limiting magnitude $V$ = 23 is shown here to
  illustrate the strong concentration of these populations towards the
  Galactic Plane. In the M$_V$ = 5 panel the extra sources at $d$ = 8kpc
  are caused by the Galactic Bulge.\label{fig:uvex2}}
\end{figure*}

The availability of wide field CCD camera arrays on medium-sized
telescopes has led to the emergence of large scale optical surveys of
the sky, targeting one or more scientific objectives. The vast
majority of these surveys, such as the Sloan Digital Sky Survey (York
et al. 2001) are concentrated on the extragalactic universe. However,
there is an increasing interest in a more detailed study of our own
Milky Way Galaxy as well. The recently completed SDSS-II/SEGUE program
(Yanny, Rockosi, Newberg et al., 2009) is specifically targeting lower
galactic latitudes, and the RAVE survey (Steinmetz et al., 2006) is
targeting the Galactic halo and streams and companions to the Milky
Way.

\begin{figure*}
\centerline{\epsfig{figure=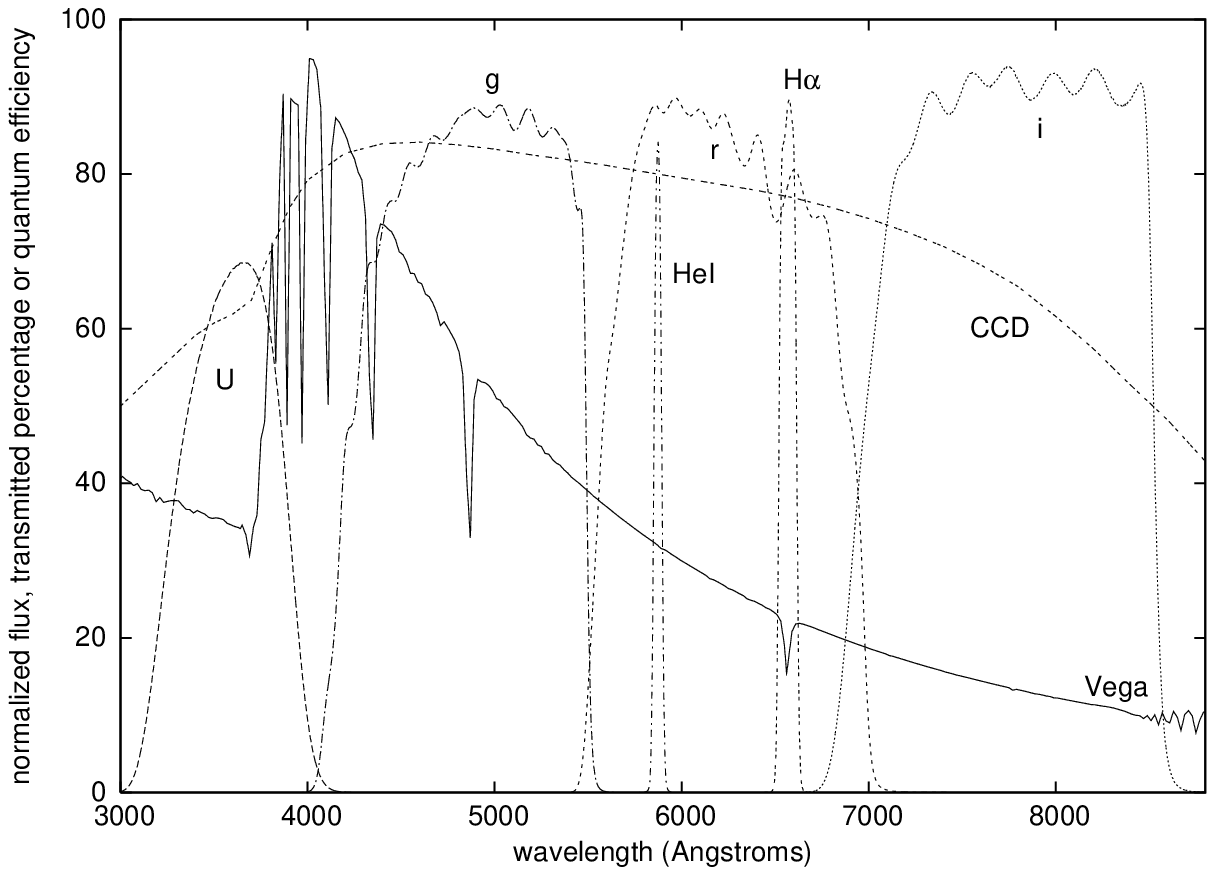,width=14cm,angle=0}}
\caption[]{Filter efficiency curves of the $U$, $g$, $r$, $He${\sc i}$5875$,
   $H\alpha$ and $i$ band filters used in the \UVEX and \IPHAS surveys
   (dashed and dashed-dotted curves),
  overplotted onto the spectrum of Vega (solid curve), together with the CCD
  efficiency curve (dashed). \label{fig:filters}}
\end{figure*}

\medskip The plane of the Milky Way, however, has so far not been the
target of a full scale optical, multicolour, digital and photon-noise
limited survey, despite its obvious value for many fields of
astrophysics. This situation is changing with the European Galactic
Plane Surveys (EGAPS) currently underway. Combined, the EGAPS surveys
(described below) will cover the full Galactic Plane in a strip of
10$\times$360 degrees centred on the Galactic equator, in the $u/U,
g, r, i$, $H\alpha$ bands down to (Vega) magnitude $\sim$21 - 22 (or
equivalent line flux), and for the Northern survey also in the \mbox{$He${\sc
  i}$5875$} filter.  EGAPS started off with the INT Photometric
H$\alpha$ Survey (\IPHAS; Drew et al., 2005, hereafter D05) that
covers the Northern Galactic Plane in the $r, i$ and $H\alpha$ bands.
Here we describe the \UVEX survey: the UV-Excess Survey of the Northern
Galactic Plane that uses the exact same set-up as \IPHAS but will image
the Northern Plane in $U,g,r$ and \mbox{$He${\sc i}$5875$}.  The
Southern Galactic Plane will be covered by the {\sl VPHAS+} survey (using
$u,g,r,i,H\alpha$) as an ESO Public Survey on the VST+Omegacam
combination and will probably start in the beginning of 2010.

\medskip Very few dedicated blue surveys at low galactic latitudes
exist, apart from all-sky surveys such as the Palomar sky surveys or
the ESO sky surveys, performed in the 1950-1990s, using photographic
plates. A notable exception is the Sandage Two-Color survey of the
Galactic Plane as presented in a series of papers by Lanning (1973)
and Lanning \& Meakes (2004; and references therein; also see Lanning
\& L\'epine, 2006). The total survey covers 5332 square degrees (124
plates) centered on the Galactic Plane (at \mbox{$b$ = --6\degr, 0\degr and
+6\degr}) using photographic plates (6.6 degrees on a side) and the UG1
(`UV') and GG13 (`B') filters on the Palomar 48-inch Oschin Schmidt
telescope.  So far 734 UV-bright sources in 39\% of the total area of the
survey have been published, averaging one source per 2.83 square
degrees down to $m_B \sim$ 20. UV-bright candidates were selected by eye
as having $U-B < 0$, but with significant scatter on both the photometry
and the completeness due to crowding and differing quality of the
photographic plates. Not being digital, and the extra problems crowding
present to photographic observations, make the survey good for picking
out the bluest objects, but not useful for a systematic study of
stellar populations in the Galactic Plane.  Although \UVEX will cover a
smaller area (1850 square degrees for the Northern Survey; 3600 square
degrees when ultimately combined with {\sl VPHAS+}), the survey is
fully photometric, allowing a more consistent and more detailed
source extraction and all data will be digitally available. First
results on the selection of UV-bright sources in \UVEX shows a much
higher surface density of sources ($\sim$10 per square degree), which
is both due to the greater depth of \UVEX (in particular in the
$g$-band), as well as the more consistent and automated selection
techniques, which allow identification of UV-bright objects at a much
redder cut-off than was possible for the Sandage Two-Color survey (see
Groot et al., 2009, {\sl in preparation}).

\medskip
Here we describe the scientific objectives of the \UVEX survey
(Sect.\ \ref{sec:science}), the survey design and observing strategy
(Sect.\ \ref{sec:survey}), simulations of stellar populations in the
\UVEX colours (Sect.\ \ref{sec:simu}), the early results
(Sect.\ \ref{sec:results}), the seeing statistics and the effect of
crowding on the detected number counts (Sect.\ \ref{sec:seeing}) and
conclusions (Sect.\ \ref{sec:conclusions}).

\section{Science goals} \label{sec:science}

\begin{figure}
\centerline{\psfig{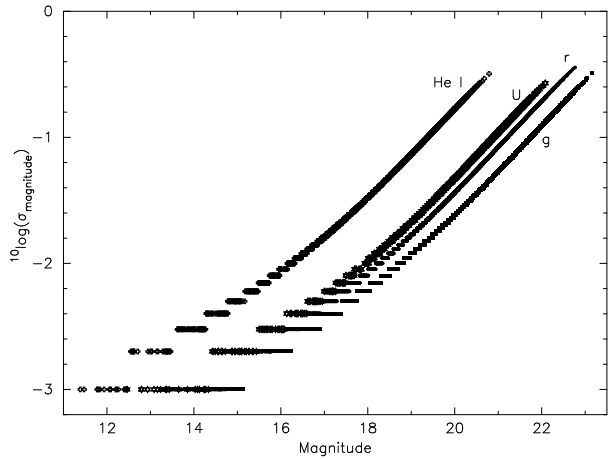}}
\caption[]{Magnitude errors for the observations in the four different
  filters as function of magnitude, as detected in CCD4 of field 6160
  taken under median seeing conditions. Note that the He{\sc i} data
  is applicable for the 120 second integrations. After the first year
  the depth in He{\sc i} is increaded by an enlarged integration time
  (180 seconds). 
\label{fig:sigmamag}}
\end{figure}

The main science goal of \UVEX is to chart the Galactic population of
stellar remnants, single and in binary systems. These include single
and binary white dwarfs, subdwarf B stars, Cataclysmic Variables, AM
CVn stars and neutron star and black hole binaries. These systems are
hot, and therefore blue, due to the remnant energy in the compact
objects or they are being kept hot due to accretion. Due to their
small size (typically $\lsim$1\,$R_\odot$) these systems are
intrinscically faint, despite their hot temperature. In particular,
they have much lower absolute visual magnitudes than main sequence
stars of similar colours. At a given apparent magnitude they will
therefore be much closer by than main sequence stars and therefore
have suffered much less extinction than a main sequence star of the
same intrinsic colour. This technique to identify stellar remnants has
been used before, e.g. to search for old halo white dwarfs in front of
molecular clouds (Hodgkin et al. {\sl priv.com.}). The reason to
survey the Galactic Plane is that the target populations are Galactic
populations and therefore strongly concentrated towards the Galactic
Plane. Fig.\ \ref{fig:uvex2} illustrates this point. Here we have
taken a model of the Galaxy according to the prescription of Boissier
\& Prantzos (1999), and populated this with populations having
absolute visual magnitudes of M$_V$ = 15, 10 and 5.  A Sandage (1972)
type model of Galactic extinction was included. This Galaxy model is
identical to the one used and described more extensively in Nelemans,
Yungelson \& Portegies Zwart (2004).  A limiting magnitude of $V$ = 23
was taken to construct Fig.\ \ref{fig:uvex2}. It can be seen that any
population with an absolute magnitude in the range 5$\leq M_V \leq $10
is strongly concentrated to the Plane of the Galaxy.  Respectively
12\%, 40\% and 97\% of all objects in Fig.\ \ref{fig:uvex2} lie within
the first 5$\degr$ of the Plane (the limits of the \UVEX survey) for
$M_V$ = 15, 10 and 5. Subdwarf B stars, Cataclysmic Variables, AM CVn
stars, young white dwarfs and most neutron star and black hole
binaries all have absolute magnitudes $M_V <$ 15.  It is only for the
faintest systems (old white dwarfs and very low mass accretion rate
interacting binaries) that we sample such a local population that
no concentration towards the Galactic plane is seen. 

\begin{figure*}
\centerline{\epsfig{figure=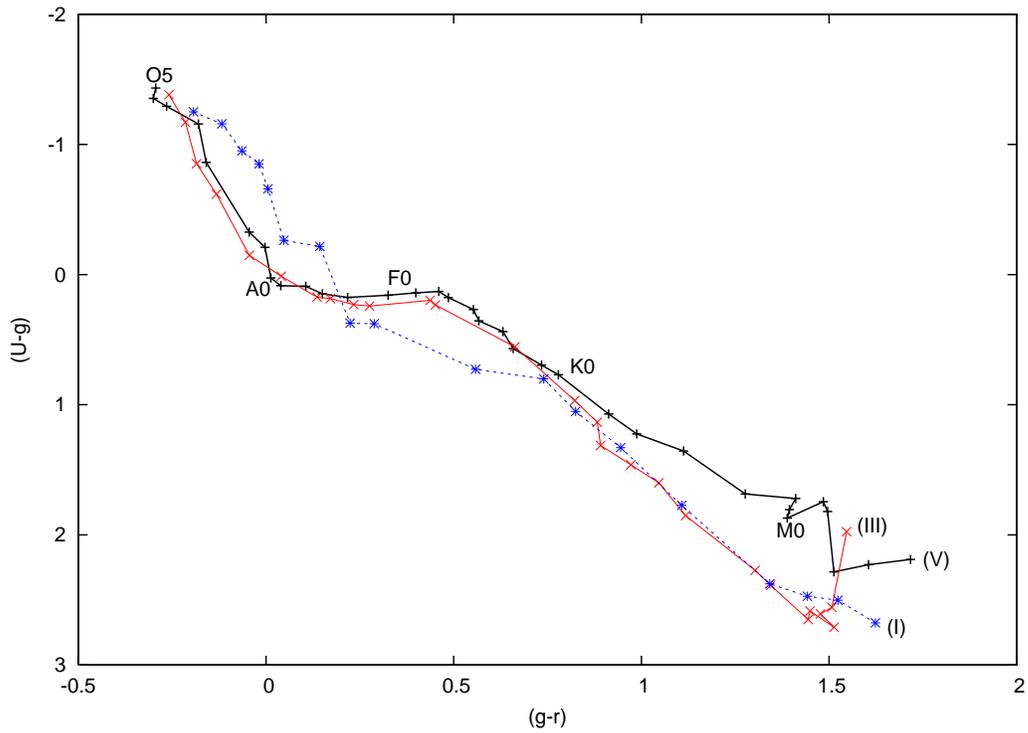,width=14cm}}
\caption[]{Simulated colours of main sequence (V,black), giant (III, red) and
  supergiant (I, blue) stars in the \UVEX $(U-g)$ vs $(g-r)$
  plane. \label{fig:uggr}}
\end{figure*} 

\begin{figure*}
\centerline{\epsfig{figure=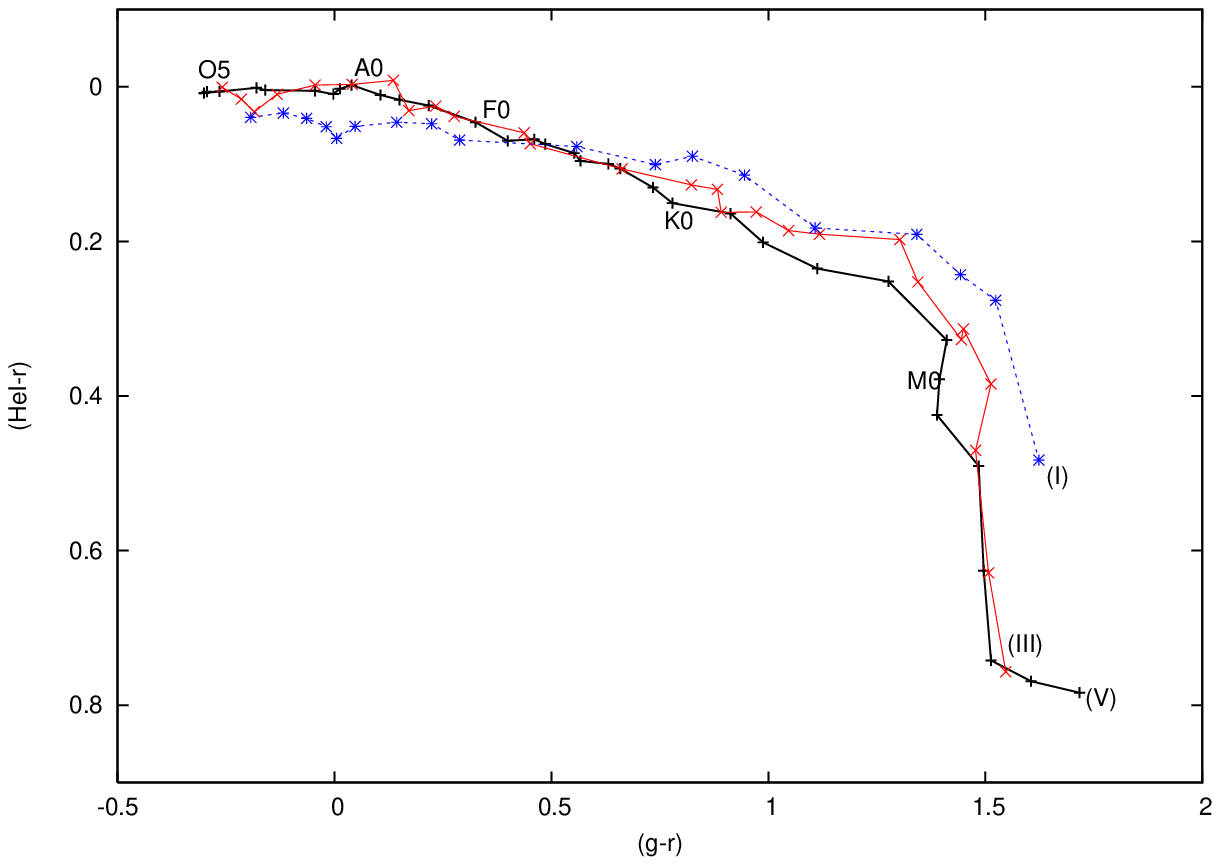,width=14cm}}
\caption[]{Simulated colours of main sequence (V, black), giant (III, red) and
  supergiant (I, blue) stars in the \UVEX (\heir) vs $(g-r)$
  plane. \label{fig:hergr}}
\end{figure*} 

\medskip The prime motivation to chart the population of interacting
binaries and stellar remnants in our Galaxy is that a large and
homogeneous sample is needed to answer questions in the fields of
binary stellar evolution (e.g. on the physics of the common-envelope
phase), the gravitational radiation foreground from compact binaries
in our Galaxy for missions such as {\sl LISA}, and the influence of
chemical composition on accretion disk physics. For this last item in
particular the comparison between hydrogen-rich systems such as
Cataclysmic variables and helium-rich (AM CVn stars; e.g. Roelofs et
al., 2006) or even C/O-rich (Ultracompact X-ray Binaries; Nelemans et
al. 2004) systems will be important. The currently known populations
of these last two classes are limited to less than two dozen systems
each, severely limiting a population study (see Roelofs, Nelemans \&
Groot, 2007).

\begin{figure}
\centerline{\epsfig{figure=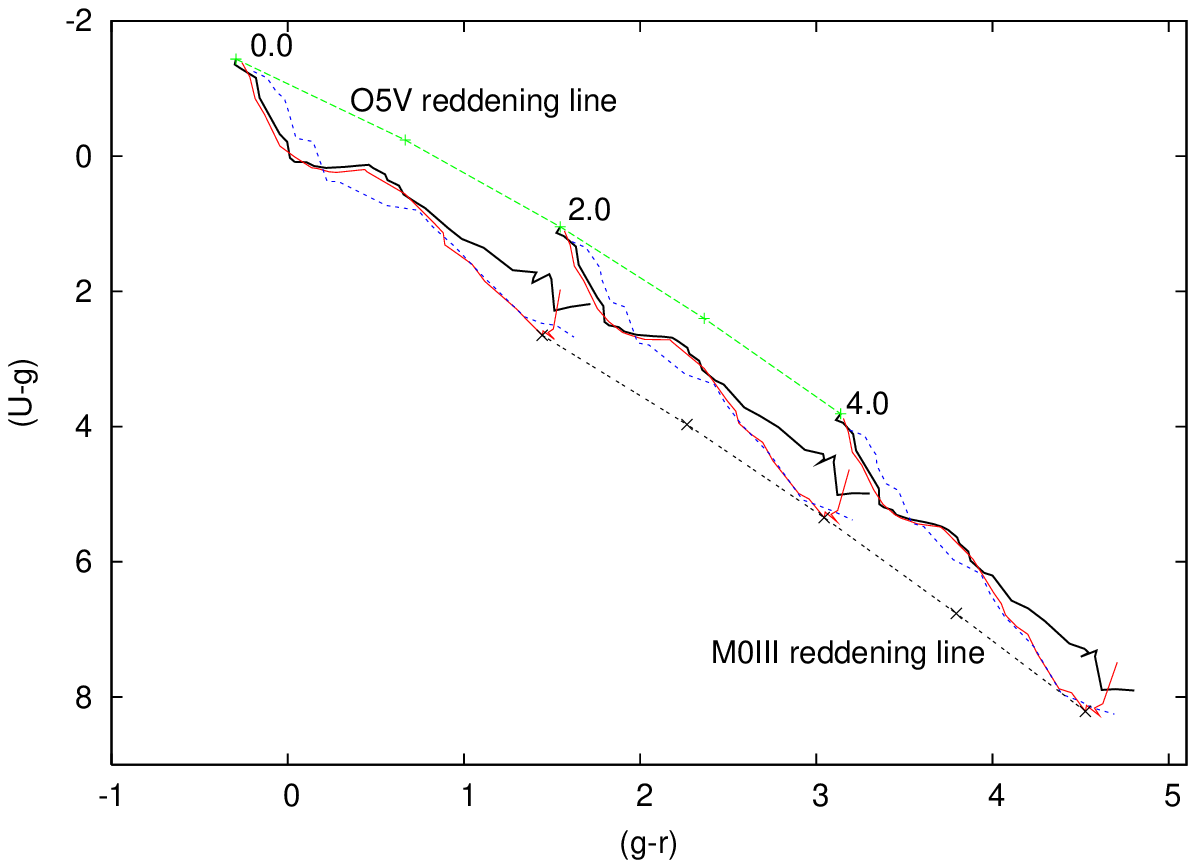,width=8.8cm}}
\centerline{\epsfig{figure=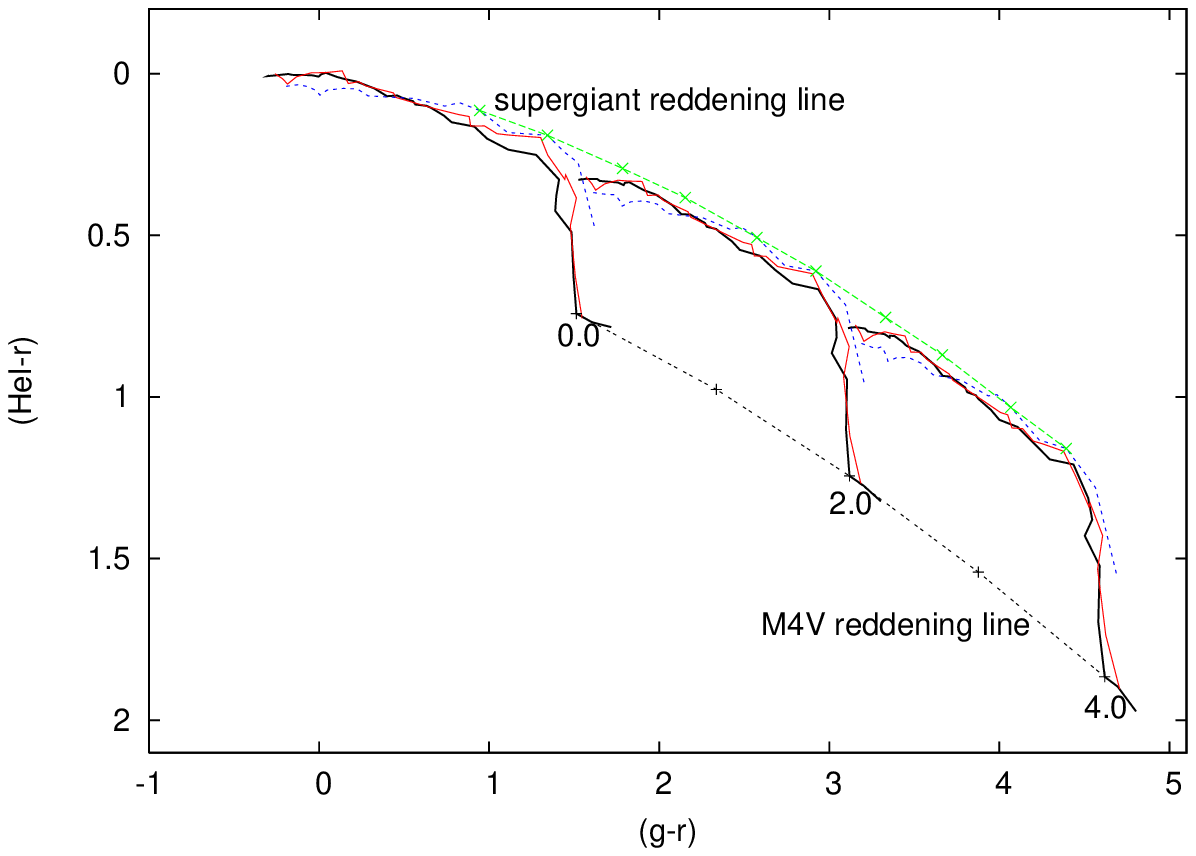,width=9.0cm}}
\caption[]{Synthetic colour-colour diagrams in the \UVEX filter system
  for main-sequences, giant and supergiant, using the reddening law of
  Cardelli et al. (1989). Colours are shown for $E(B-V) = 0.0, 2.0$ and
  $4.0$. Also shown are the encompassing upper and lower envelope
  curves: the O5V reddening curve and the M0{\sc iii} reddening
  curve. \label{fig:reduvex}}
\end{figure}

\medskip
Besides the main science goal outlined above, the {\sl UVEX} survey
will allow for many more scientific studies in the field of Galactic
astrophysics, especially in combination with the {\sl IPHAS}
survey. The combination of {\sl UVEX} and {\sl IPHAS} will allow a
much cleaner separation of stellar populations with different
intrinsic colours, absolute magnitudes and varying degrees of
reddening. The added opportunities include:
\begin{itemize}
\item A 3D dust model of our Galaxy on an arcsecond scale: the
  inclusion of H$\alpha$ with the four broad bands allows for breaking
  the degeneracy between reddened early type stars and unreddened late
  type stars due to their strongly different H$\alpha$ absorption
  lines strength. See Drew et al. (2008) and Sale et al. (2009) for the
  usage of \IPHAS data towards this goal. 
\item The first ever accurate high proper motion study in the Galactic
  Plane. The overlap in $r$ band observations between {\sl
  UVEX} and {\sl IPHAS} has a minimum of three years baseline distance,
  and otherwise identical set-up allows for proper motion determinations
  down to $\mu\geq$100 mas yr$^{-1}$ for the {\sl IPHAS - UVEX}
  comparison and down to $\mu\geq$20 mas yr$^{-1}$ for an {\sl IPHAS -
    POSS-I} comparison; Deacon et al. (2009). These should be compared
  to the recent surveys by L\'epine et al. (see L\'epine 2008 and L\'epine
  \& Shara, 2005). The combined {\sl IPHAS/UVEX} proper motions
  will have more accurate photometry and better completeness, both
  because CCD observations are used instead of photographic plates.
\item The identification and characterisation of open clusters, star
  forming regions and (highly reddened) O/B associations down to
  the magnitude limit of the two surveys. 
\item The characterization of stellar photometric variability on a three-year
  timescale due to reobservations in the $r$ band. This
  includes the identification of long period variables (e.g. Mira
  stars), irregular variables (dwarf novae-type cataclysmic variables,
  soft X-ray transients, flare stars), and a first identification of
  large amplitude regular variables (e.g. RR Lyrae stars). 
{\sl EGAPS} will also serve as a baseline for new transients in the
plane of the Milky Way as shown by our study of Nova Vul 2007 (V458
Vul) whose progenitor was identified in \IPHAS data taken only seven
weeks before the nova explosion (Wesson et al., 2008).    
\end{itemize}

In the science goals of the \UVEX survey the $U$-band observations play
a crucial role. Since the $U$-band is the most sensitive to dust
extinction it is not only a pivotal band to identify those hot but
low-luminosity populations in the Plane, but is also a band that will
play a very important role in the dust mapping of the Plane. Straddling the
Balmer jump it is also the broad band which is most sensitive to
chemical composition and atmospheric pressure in the underlying populations. 

\section{Survey Design \& Data processing}
\label{sec:survey}

Apart from the filters, the survey design is identical to that of the
{\sl IPHAS} survey as described in D05. The same field centers have
been taken so all fields will be imaged twice in a set of overlapping
pointings. The order of field selection for {\sl UVEX} is mainly based
on the availability of `good' {\sl IPHAS} data for the same field with
a time baseline of at least three years. Here `good' {\sl IPHAS}
photometry refers to those fields that have a seeing less than
2.0\arcsec, ellipticity of the stellar image $<$0.2 and a sky
background of $<$2000 cts in $r$ (Gonz{\'a}lez-Solares et al., 2008).
This strategy has been chosen to ensure both a reasonable proper
motion baseline as well as a high quality dataset covering the full
optical spectrum. {\sl UVEX} observations are done in the RGO $U$
filter, the Sloan Gunn $g$ and $r$ filters and the 
\mbox{$He${\sc i}$5875$} filter.  Integration times are 120 sec ($U$), 30 sec
($g$), 30 sec ($r$) and 120/180 sec ($He${\sc i}).

Fig.\ \ref{fig:filters} shows the throughput of the $U,g,r$ and
\mbox{$He${\sc i}$5875$} filters, as well as the \IPHAS $r$, $i$ and $H\alpha$,
overplotted onto the spectrum of Vega.  As can be seen in Fig.\
\ref{fig:filters} the \mbox{$He${\sc i}$5875$} filter overlaps with the
$r$-band filter, but has a slightly bluer effective wavelength than
the $r$-band. For this reason we construct the (\heir) colour,
adhering to the usual notation for colours to list the bluer band
first. Note that the $r$-band curve is slightly
different from that shown in D05, even though it is the
same filter. Filter efficiencies were remeasured in July 2006 at the
ING Observatory, resulting in the current efficiency curves. 

Data processing is also identical to the \IPHAS procedure. All data is
transported from the telescope to the Cambridge Astronomy Survey Unit,
where it is processed according to the pipeline procedure as detailed
in Irwin \& Lewis (2001), D05 and Gonz\'{a}lez-Solares et al.(2008).


All magnitudes are on the Vega system. The $g$ and $r$
band observations are calibrated on a nightly basis by the observation
of photometric standards stars from Landolt (1992).  Photometric
calibration in $U$, $H\alpha$ and \mbox{$He${\sc i}$5875$} is hard-coupled to
that of the $g$-band (for $U$), and the $r$-band filter (for $H\alpha$
and \mbox{$He${\sc i}$5875$}). The fixed offsets (in Vega magnitudes) are $U =
g-2.100$, $H\alpha = r-3.140$ and \mbox{$He${\sc i}} = $r-3.575$. These
shifts have been determined on the basis of spectrophotometric
observations combined with the colour-modeling discussed in Section\
\ref{sec:simu}. On a typical good night the zeropoints (in ADU) for
$g$ and $r$ are 25.01 and 24.51 respectively, showing the greater
depth of the $g$-band observations for a given integration time. Since
the $g$- and $r$-band observations are both 30 seconds, the $g$-band
gives the deepest observations of the combined \IPHAS/\UVEX survey. A
global photometric calibration of both surveys remains to be done.

For illustrative purposes Fig.\ \ref{fig:sigmamag} shows the magnitude
error as a function of magnitude and colour for the \UVEX observations
of field 6160, where we have taken all data which have been marked as
`stellar' (quality flag `--1' as defined in Gonz\'{a}lez-Solares et
al., 2008) and on CCD 4 of the Wide Field Camera.
Fig.\ \ref{fig:uvex6160} shows the colour-magnitude and colour-colour
diagrams for the same field.

\section{Simulation of the UVEX colour-colour planes}
\label{sec:simu}

To interpret the \UVEX observations, simulations of the colours of
stars and the effect of reddening are a very powerful and important
tool. In obtaining the simulated colours we follow the procedure as
outlined in D05 for the \IPHAS survey. In short, model and template
spectra are folded with the efficiency curves as shown in
Fig.\ \ref{fig:filters} and the CCD response curve, and calibrated on
the Vega system using Eq.  1 of D05, where $r$ and $i$ indices should
be replaced with the appropriate filter curves. The only difference in
this procedure is that we did not rebin all input data to a
5\AA\ resolution (in D05 set by the resolution of the Pickles, 1998,
library), but used a fixed 1\AA\ sampling and a linear interpolation
where necessary, as well as an extrapolation on the CCD-efficiency
curve on the blue-side of the $U$-band since data was not available. A
check on the colours obtained has been made by reproducing the colours
as given in D05 for the \IPHAS filters (using the filter curves as
given in D05). The mean difference and standard deviation on the \IPHAS
colours derived in D05 and here are $\overline{\Delta(r-i)} = 0.009$
and $\sigma_{(r-i)} = 0.002$ and $\overline{\Delta(r-H\alpha)} =
0.006$ and $\sigma_{(r-H\alpha)} = 0.002$.  Further checks to the
procedure were made by inserting Johnson-Cousins filters into the
equation and calculating the colours of main-sequence stars, based on
the Pickles spectra, and compared with the colours as given in Bessell
(1990) for main-sequence and giant stars and with the Stone \& Baldwin
(1983) southern spectrophotometric standard stars as given by Landolt
(1992).

Our conclusion from these comparisons is that the method accurately
reproduces the colours of stars as given in the literature, although
there is a large scatter on the $(U-B)$ colours.  The relatively large
scatter with respect to the stars given in Bessell (1990) can also be
attributed to the use of a different set of input spectra (the Vilnius
spectra used by Bessell vs. the Pickles spectra used by us). The
comparison with the Baldwin-Stone spectrophotometric standards as
given by Landolt (1992) and the colours from D05 show the accuracy of
the method. All synthetic colours calculated in this paper are given
in the Appendices. The large scatter in the U-band is not surprising
given its sensitivity to metallicity, atmospheric absorption, and the
strong variations in detector reponses that occur at the bluest
wavelengths.  This procedure was then used to derive the colours in
the \UVEX filters, for normal stars (luminosity classes V, III and I),
emission line objects and white dwarfs.

\medskip
Based on the colour-simulations presented below, and also folding the
Pickles spectra with standard Johnson-Cousins filter curves we derive
the following colour-transformations from Johnson-Cousins to the
\IPHAS/\UVEX colour space: 

\begin{eqnarray}
(U-g) = 1.035\,(U-B) + 0.470\,(B-V) - 0.017\\
(g-r) = 1.044\,(B-V) - 0.116\,(V-R) + 0.025\\
(r-i) = 1.484\,(R-I) - 0.389\,(V-R) - 0.014
\end{eqnarray}

These transformations are valid over the colour-region of
$-1.43<(U-g)<2.19$, $-0.30<(g-r)<1.72$ and
$-0.17<(r-i)<2.58$. Please note that all magnitudes here are on the
Vega system. Transferring to the AB system can be done by using the
relations given on the Astronomy Survey Unit's webpage\footnote{
\mbox{http://www.ast.cam.ac.uk/$\sim$wfcsur/technical/photom/colours/}}.

\begin{figure}
\centerline{\epsfig{figure=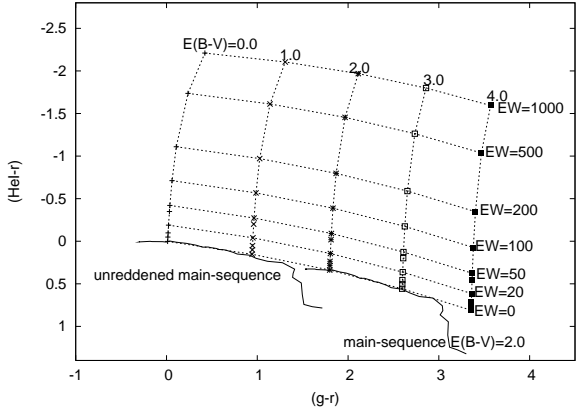,width=8cm}}
\caption[]{Position of \mbox{He{\sc i}$5875$} emission line objects in the
  $(g-r)$ vs. (\heir) colour space. Horizontal lines mark objects of equal
  equivalent widths (EW) in units of \AA, vertical lines mark lines of
  constant reddening, assuming an underlying A0V spectrum.
\label{fig:heemission}}
\end{figure}

\subsection{\UVEX colours of main sequence, giant and supergiant stars}
\label{sec:normalcolours}

To simulate the colours of normal stars with luminosity classes V
(main-sequence dwarfs), III (giants) and I (supergiants) we make use
of the Pickles (1998) library. After application of the procedure
outlined above, the results are shown in Figs.\, \ref{fig:uggr} \&
\ref{fig:hergr} and are tabulated in Appendix\, \ref{app:uvexcolours}.
Here we use the $(g-r)$ vs. $(U-g)$ and the $(g-r)$ vs. (\heir)
colour-colour diagrams as our fundamental planes. It can be seen that
the difference in colours between main-sequence stars, giants and
supergiants is relatively small and all objects are restricted to a
narrow band. In the ($g-r$) vs. (\heir) plane a characteristic
`hook' is displayed, appearing around M0, after which the stars show a
distinct increase in the (\heir) colour, caused by the appearance of
strong TiO absorption bands depressing the flux in the He{\sc i}
band.

To simulate the effect of reddening we have applied the extinction
laws of Cardelli, Clayton \& Mathis (1989), with a fixed $R$=3.1.
Results are shown in Fig.\ \ref{fig:reduvex}. Template and model
spectra were first multiplied by the extinction laws and then folded
through the filter curves.  It is clear from Fig.\ \ref{fig:reduvex}
that, in analogy to the \IPHAS colours, also here envelope lines
exist, indicating a limit above/underneath which no normal
main-sequence stars, giants or supergiants are expected. In
Fig.\ \ref{fig:reduvex} these are indicated with `O5V-reddening' line
and `M0III reddening line'.

\begin{figure}
\centerline{\epsfig{figure=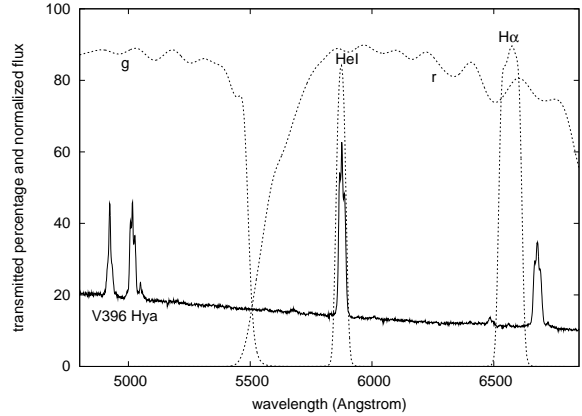,width=8cm}}
\caption[]{Spectrum of the AM CVn star V396 Hya (Ruiz et
  al., 2001), overplotted with the \UVEX/\IPHAS narrow-band filter
  curves of \mbox{$He${\sc i}$5875$} and $H\alpha$ and the broad-band
  $g$ and $r$ bands. 
\label{fig:specamcvn}}
\end{figure} 

\begin{figure}
\centerline{\epsfig{figure=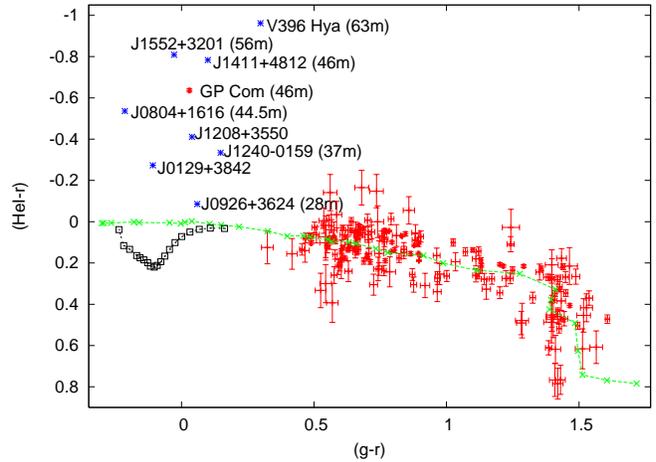,width=9cm,angle=0}}
\caption[]{ Position of AM CVn stars in a ($g-r$) vs. (\heir)
  colour-colour diagram. Asterisks show the position of known AM CVn
  stars, based on their SDSS spectrum. Labels refer to the AM CVn
  stars and, if known, the orbital period is added (see Roelofs et
  al., 2009 for SDSS\,J0804). The AM CVn points are overlayed onto the
  \UVEX colours of stars in the field of GP Com (red symbols), which
  itself is shown by the (red) point at $(g-r) = 0.05$ \& $(\heir) =
  -0.65$. Also shown is the unreddened main-sequence as the green
  dashed line marked by crosses, and the DB sequence as the purple
  dotted line marked with squares.
\label{fig:gpcom_amcvn}}
\end{figure}

\subsection{Colours of  helium emission-line stars}
\label{sec:emission}

The inclusion of the \mbox{$He${\sc i}$5875$} filter has been made to enable
the detection of strong \mbox{$He${\sc i}$5875$} absorbers (e.g. DB type white
dwarfs) or emitters (e.g. AM Canum Venaticorum stars and Cataclysmic
Variables). Again, in analogy with D05 we have determined the
sensitivity to pick out He{\sc i} emission using an A0V underlying
continuum (very similar to a power law slope with index --3) to which
an emission line is added. The emission line is simulated by a 
top-hat shaped line having a width that is equal to the
full-width-at-half-maximum of the He{\sc i} filter, 40\AA. The
results are shown in Fig.\ \ref{fig:heemission}. Reddening has been
added to these data in discrete steps of 1 from $E(B-V) = 0$ to $E(B-V)=4$.
Overplotted onto the grid of emission line strength is the unreddened
main-sequence as determined in Sect.\ \ref{sec:normalcolours}. It can
be seen from Fig\ \ref{fig:heemission} that emission strength above
already a few \AA\ should stand out in \UVEX observations, depending
mostly on the photometric accuracy of the observations. This is only
marginally influenced by reddening due to the narrowness of the He{\sc
  i} filter and its position within the $r$-band. A similar behaviour
is seen in the $(r-H\alpha$) colour although the effect is larger
there (D05). 


\subsection{The colours of AM Canum Venaticorum stars}
\label{sec:amcvn}

AM Canum Venaticorum (AM CVn) stars are hydrogen-depleted,
short-period interacting binaries consisting of a white dwarf primary
and a white dwarf or semi-degenerate helium star secondary, sometimes
also called `Helium Cataclysmic Variables' (see e.g. Nelemans, 2005,
for an overview). These systems show orbital periods in the range 5.4
min $<P_{\rm orb} < 65$ min.  At longer orbital periods ($P_{\rm orb}
\gsim 30$ min) their spectra are dominated by strong helium emission
lines (see e.g.  Marsh, 1991; Roelofs et al. 2005,2006a,b). The
strongest of these lines is the \mbox{He{\sc i}5875}
line. Fig.\ \ref{fig:specamcvn} shows the spectrum of V396 Hya (Ruiz
et al., 2001) with the \UVEX/\IPHAS narrow-band filters of
\mbox{$He${\sc i}$5875$} and H$\alpha$ overplotted. It can be seen
that the \mbox{$He${\sc i}$5875$} filter width exactly matches the
width of the emission line and therefore provides maximum sensitivity
to these systems.

Using the publicly available Sloan spectra and the Very Large
Telescope spectra as presented in Roelofs et al. (2005, 2006,
2007a,b,c) we constructed a $(g-r)$ vs. (\heir) colour-colour diagram of
long period AM CVn stars (Fig.\ \ref{fig:gpcom_amcvn}). It can be seen
that indeed the long period AM CVn stars lie significantly above the
main-sequence in (\heir) due to their \mbox{He{\sc i} 5875} emission. The
vertical spread of the AM CVn systems indicates increasing He{\sc i}
equivalent widths at almost constant broad-band colours. 


\begin{figure}
\centerline{\epsfig{figure=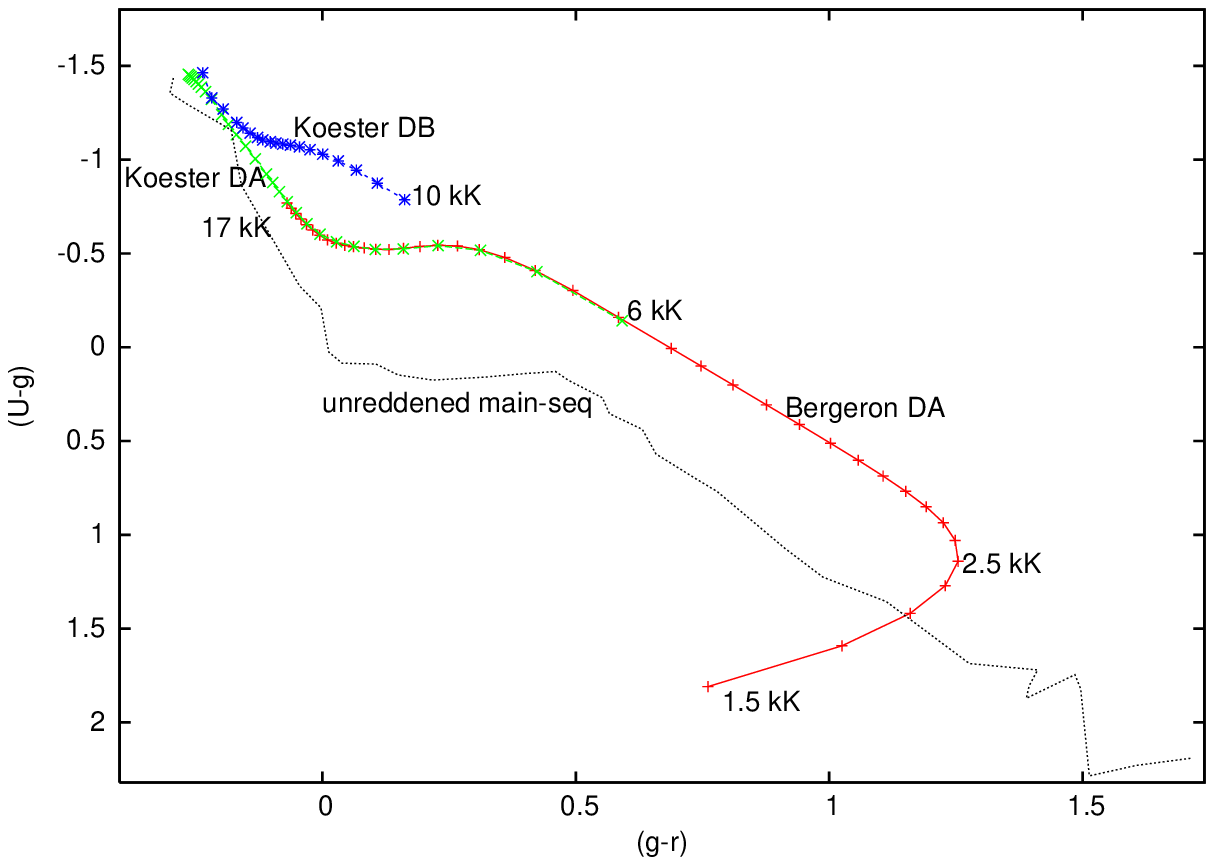,width=8.5cm}}
\centerline{\epsfig{figure=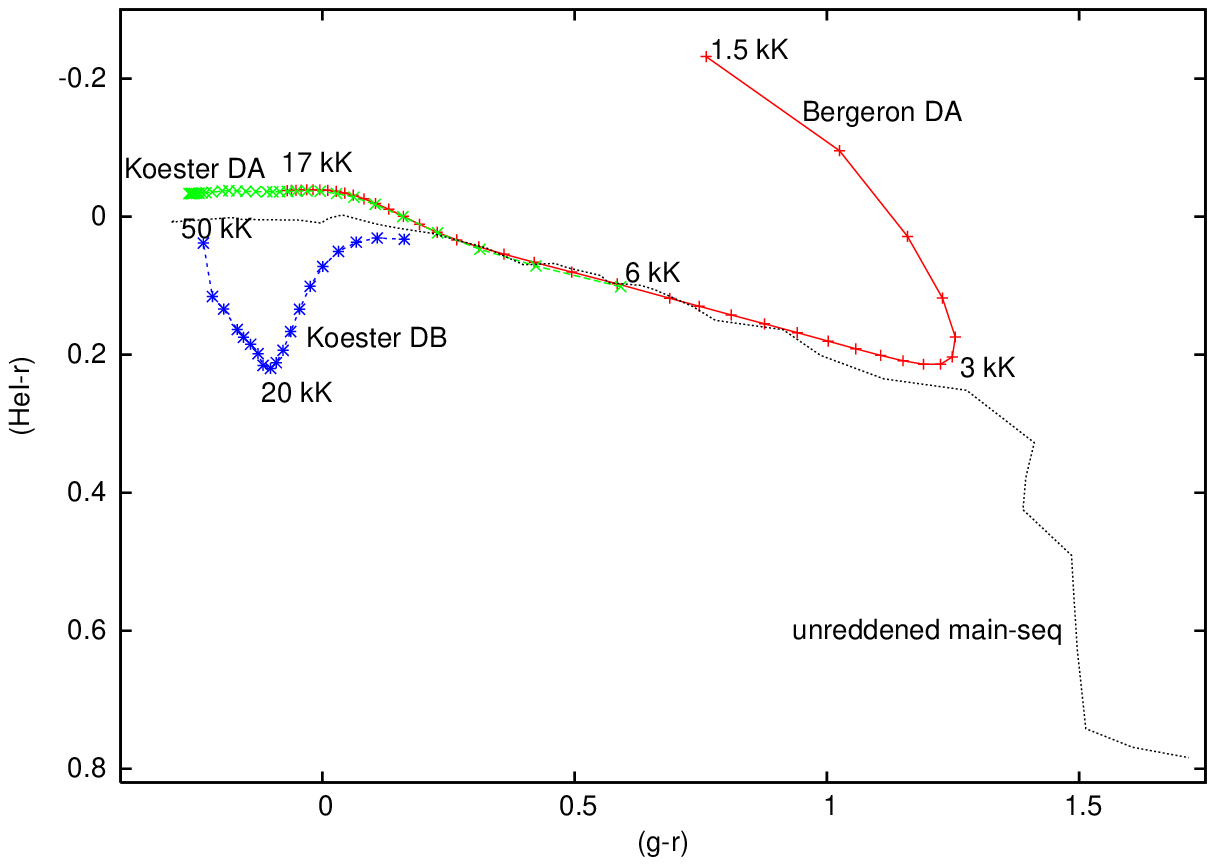,width=8.5cm}}
\centerline{\epsfig{figure=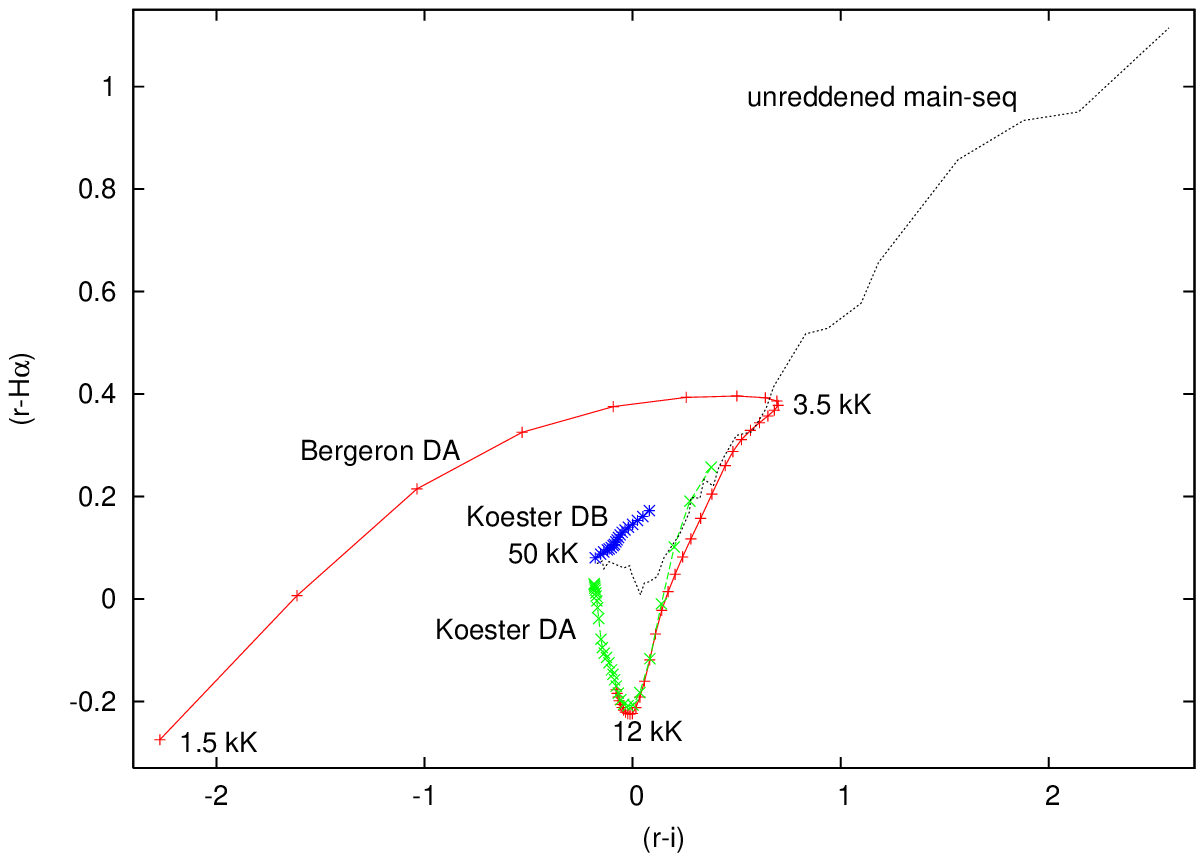,width=8.6cm}}
\caption[]{{\sl Top:} Position of $\log g$=8.0 DA and DB white dwarf
  in the \UVEX $(g-r)$ vs $(U-g)$ colour-colour plane based on the
  Bergeron models (DA) and Koester models (DA \&DB) for temperature
  between 1\,500 K and 80\,000 K (DA) and 10\,000 - 50\,000 K (DB). DB
  models with temperature $<$10,000 K are identical to
  blackbodies. The characteristic `hook' in the DA models between
  temperature 7\,000 - 25\,000 K is due to the Balmer jump which lies
  in the U-band. The hook at $T\lsim2\,500$K is due to collisionally
  induced absorption by the H$_2$ molecule. {\sl Middle:} Same models
  as in the previous panel in the ($g-r$) vs.(\heir) colour plane. The
  DA models with $T>3\,000$ K virtually overlay the main-sequence
  models shown in Fig.\ \ref{fig:hergr}. The DB models show a
  pronounced reddening of the \heir colour due to the He absorption
  line at \mbox{He{\sc i} 5875}. {\sl Bottom:} Same models as in the
  previous panels in the ($r-i$) vs. ($r-H\alpha$) colour, to show the
  distinction that can be made in these colours between DA and non-DA
  white dwarfs based on the deep H$\alpha$ absorption.
  \label{fig:wdmodels}}
\end{figure}

\subsection{Simulation of DA and DB white dwarfs}
\label{sec:whitedwarfs}

As uncovering the population of single and binary white dwarfs at low
galactic latitude is one of the main goals of the \UVEX survey we have
also simulated the expected colours of a set of white dwarfs. These
simulations are based on two sets of white dwarf model spectra
available to us: one set kindly provided by D. Koester, spanning the
temperature range 6\,000-80\,000 K and surface gravity range $\log g$
= 7.0 - 9.0, for both hydrogen dominated atmospheres (DA white dwarfs)
as well as helium dominated atmospheres (DB white dwarfs) and one set
kindly provided by P. Bergeron spanning the temperature range 1\,500 K
- 17\,000 K and surface gravity range $\log g$ = 7.0 - 9.0, for
hydrogen dominated atmospheres (DA white dwarfs). In the coolest
models (T $<$ 4\,000 K) the effect of collisional induced absorption due
to the formation of H$_2$ was included. See Finley, Koester \& Basri
(1997), Koester et al. (2001) and Bergeron, Wesemael \& Beauchamp
(1995) for details on the calculation of these models. Both sets of
models were provided on a non-linear wavelength grid, where the lines
were more densely sampled than the continuum region. Both sets of
models were interpolated on a regular grid with a 1\AA\ binning,
identical to the sampling of the filter curves and CCD efficiency. In
the overlapping region both sets of models were compared with each
other, and were found to be identical on the level of $<2\%$ at all
wavelengths with the exception of the very cores of the lines, where
differences can increase to $\sim$4\% over a small wavelength range.

\medskip
For the calculation of the white dwarf models we have used the
models with a fixed surface gravity of $\log g$ = 8.0. Fig.\
\ref{fig:wdmodels} shows the colours of the white dwarf models in the
\UVEX colour-colour planes. 

\begin{figure}
\centerline{\epsfig{figure=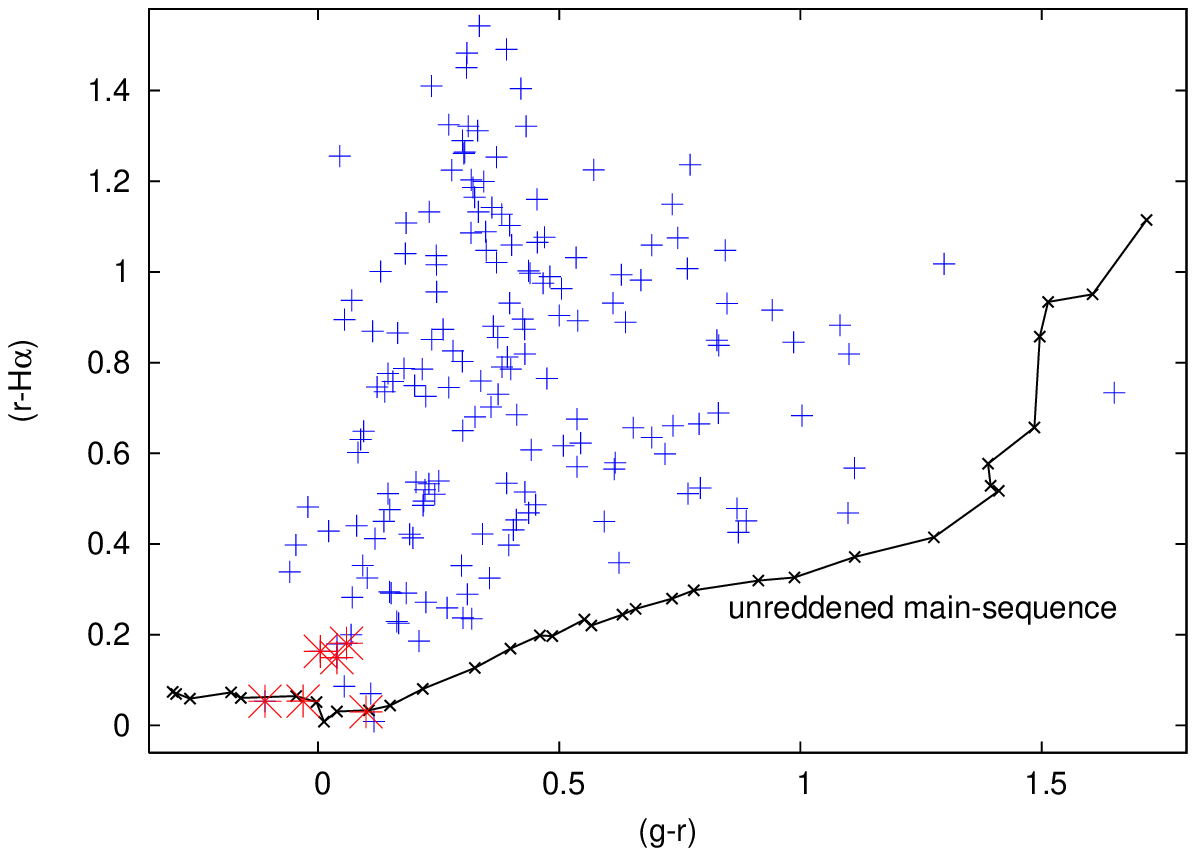,width=8cm}}
\centerline{\epsfig{figure=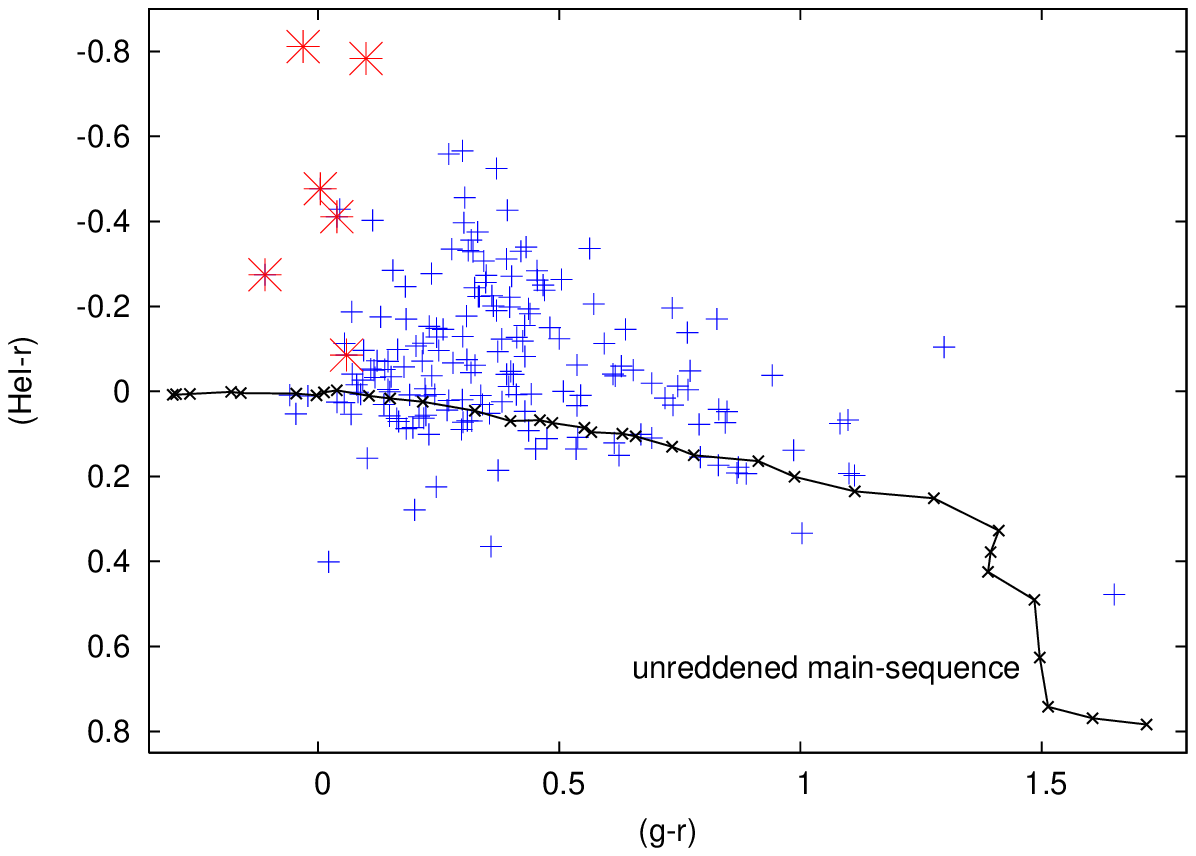,width=8cm}}
\centerline{\epsfig{figure=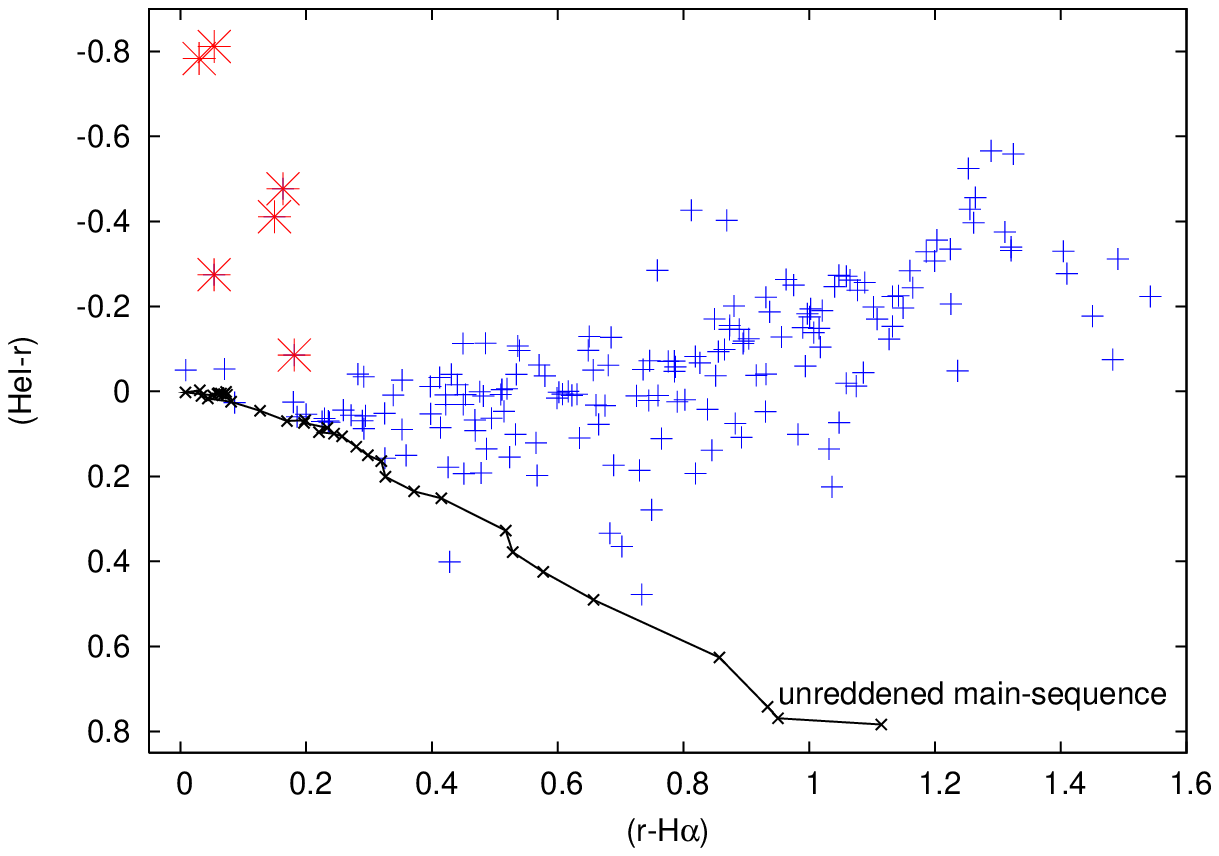,width=8cm}}
\caption[]{\UVEX/\IPHAS colour-colour diagram
  of all currently known Sloan Cataclysmic Variables (`+'-signs) and 
AM CVn stars (`*'-signs) in $(g-r)$ vs. ($R-H\alpha$) ({\sl top}),
$(g-r)$ vs. \heir ({\sl middle}) and ($r-H\alpha$) vs.(\heir) ({\sl bottom}), together with the unreddened main-sequence track.  
\label{fig:cv}}
\end{figure} 

\subsection{Simulations of Cataclysmic Variables}
\label{sec:cvs}

The location of Cataclysmic Variables in H$\alpha$ narrow-band surveys
has been extensively discussed in Witham et al., 2006. In \IPHAS, based
on solely the $r-H\alpha$ and $r-i$ colour, it is difficult to make a
photometric distinction between highly reddened background early-type
emission line objects and Cataclysmic Variables. With the addition of
the \UVEX colours this will become easier. Cataclysmic Variables are
intrinsically rather faint ($M_V\gsim$5) but blue, making them on
average much less reddened than intrinsically brighter objects at the same
colour. We have simulated the position of Cataclysmic Variables in the
\UVEX survey by taking the sample of Sloan Digital Sky Survey
Cataclysmic Variables (Szkody et al. 2002,2003,2004,2005,2006,2007)
and folded them through the \UVEX filter curves. The $U$-band magnitude
could not be calculated due to the blue cut-off in the Sloan Spectra
at $\lambda \sim$ 3800 \AA. In Fig.\ \ref{fig:cv} we show the
colour of all SDSS Cataclysmic Variables and AM
CVn stars in the \UVEX/\IPHAS colour planes. 

\begin{figure*}
\centerline{\epsfig{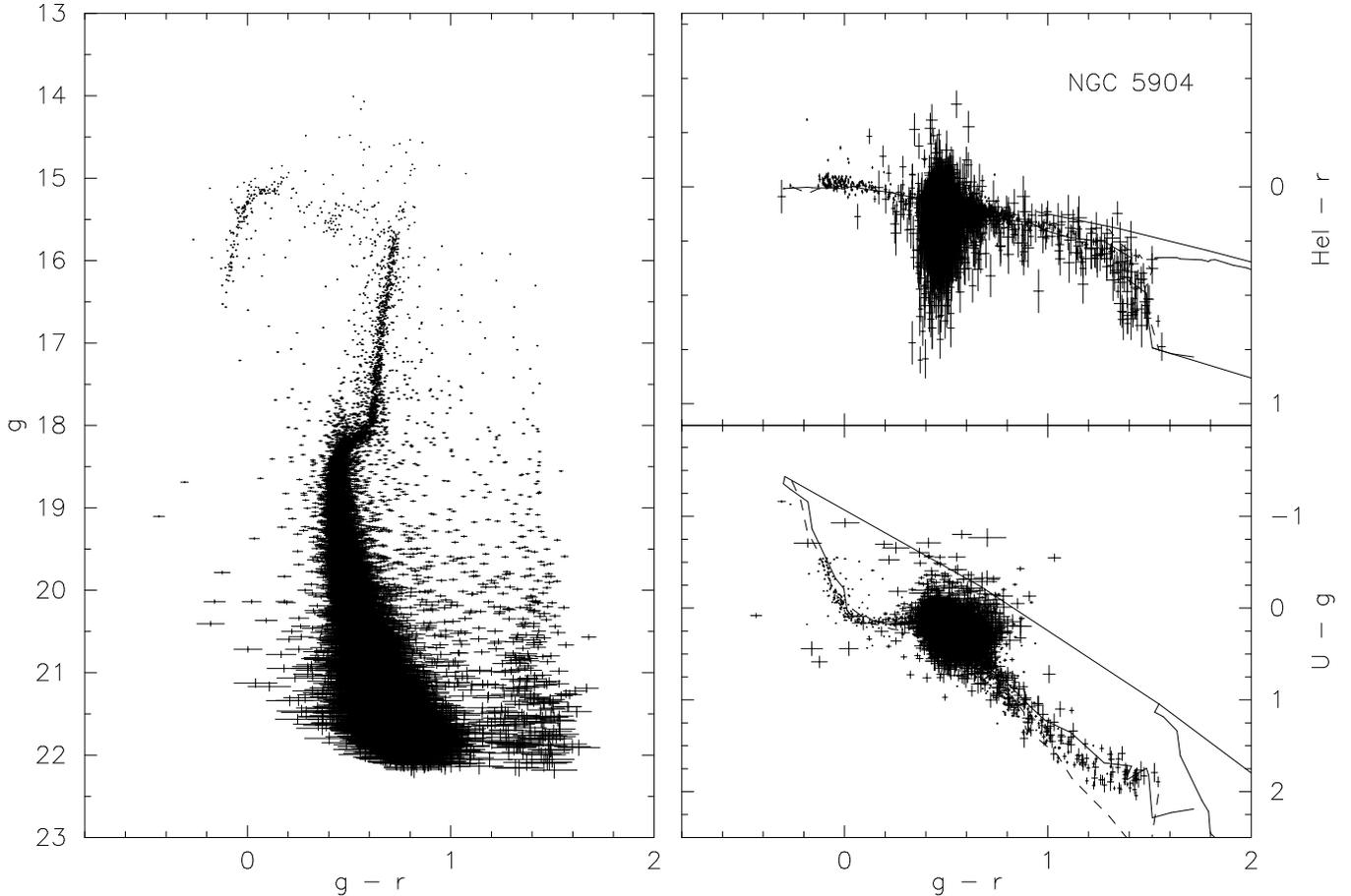}}
\caption[]{Colour-colour and colour-magnitude diagrams of globular
  cluster NGC5904 (M5), showing all detected objects with a magnitude
  error $<0.1$ for clarity, overlayed in the colour-colour diagrams
  with the \UVEX colour-tracks as presented in Sect.\ \ref{sec:simu}.
  Full lines are for main-sequence stars and dashed-lines for giants.
  The O5-reddening line and the supergiant reddening lines are also
  shown as the upper and lower envelopes (see Fig.\ \ref{fig:reduvex}
  \label{fig:ngc5904})}
\end{figure*} 

\begin{figure*}
\centerline{\epsfig{figure=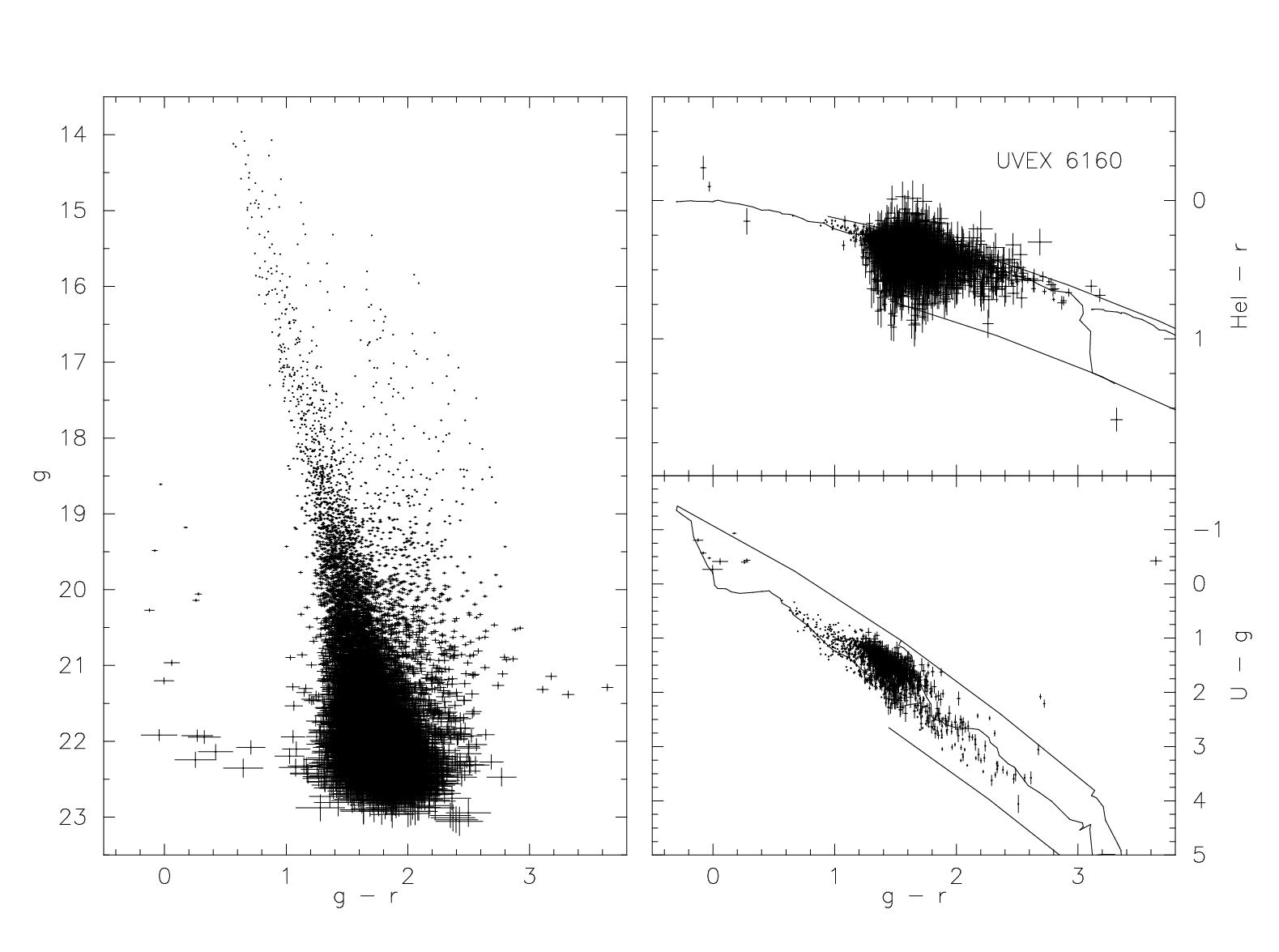,width=14cm,angle=-90}}
\caption[]{Colour-colour and colour-magnitude diagrams of field 6160, 
showing all detected objects with a  magnitude
  error $<0.1$ for clarity, overlayed in the
  colour-colour diagrams with the \UVEX colour-tracks as presented in
  Sect.\ \ref{sec:simu}
  \label{fig:uvex6160}}
\end{figure*} 

\begin{figure*}
\centerline{\epsfig{figure=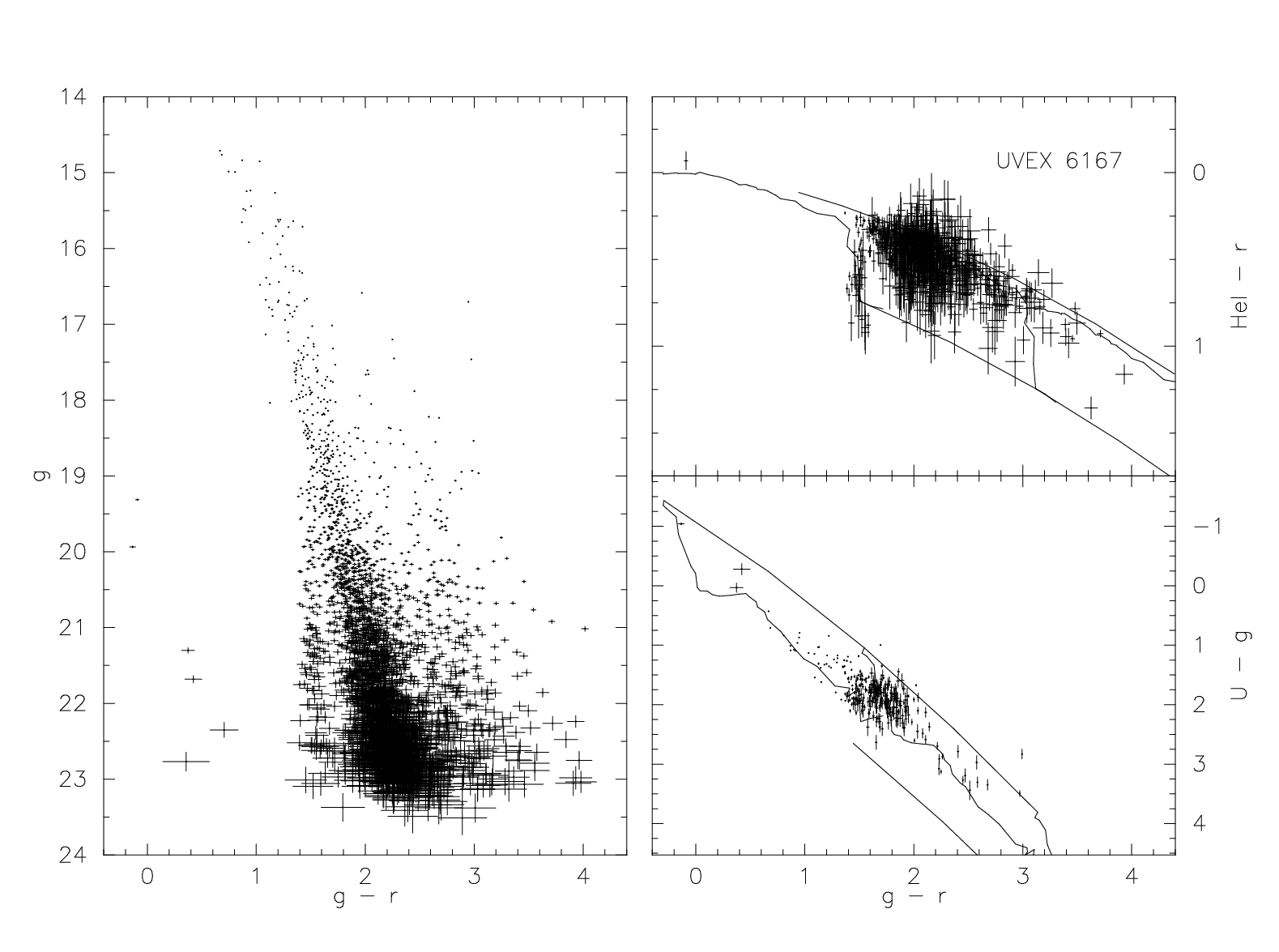,width=14cm,angle=-90}}
\caption[]{Colour-colour and colour-magnitude diagrams of field 6167, 
showing all detected objects with a  magnitude
  error $<0.1$ for clarity, overlayed in the
  colour-colour diagrams with the \UVEX colour-tracks as presented in
  Sect.\ \ref{sec:simu}
  \label{fig:uvex6167}}
\end{figure*}

\section{Comparison with observed data}
\label{sec:results}

Data taking for \UVEX has started in the summer of 2006, and up to
September 2008 30\% has been observed. After quality control checks
all data will be made public through the website of the European
Galactic Plane Surveys (EGAPS)\footnote{www.egaps.org, see also www.iphas.org}. 

\subsection{Control fields \& Survey depth}
\label{sec:gcs}

To check our photometric calibration and extraction algorithms in
highly crowded areas a number of globular clusters were observed as
control fields. In Fig.\ \ref{fig:ngc5904} we show the
colour-magnitude and colour-colour diagrams for NGC 5904, overlaid
with our main-sequence colour tracks. It is clear from
Fig.\ \ref{fig:ngc5904} that the extraction mechanism works very well,
even in severely crowded regions. The limiting magnitude (defined here
as the magnitude where the magnitude error reaches 0.2 magnitudes
(i.e. $\sim 5 \sigma$, which would be an error of 0.22 magnitudes) of
\UVEX data under good conditions ($r$-band seeing of 1\farcs1) is 21.8
($U$), 22.6 ($g$), 22.1 ($r$) and 20.2 (\mbox{$He${\sc i}$5875$}). A limiting
magnitude set at 0.2 magnitudes error in the magnitude value
encompasses between 95\% and 98\% of all stellar objects,
depending on the filter.  Part of the \mbox{$He${\sc i}$5875$} observations are
taken with 180 second integration, increasing the limiting
magnitude. In general, of course, the limiting magnitude of each
individual exposure will depend on seeing, transparency, sky
brightness and, in severe cases, also crowding (see Sect.\ \ref{sec:seeing}).

\subsection{Galactic Plane data}
\label{sec:plane}

In Figs.\ \ref{fig:uvex6160} \& \ref{fig:uvex6167} we show two
representative fields from the Galactic plane centered on
($l,b$=79.6\degr,--2.8\degr) and ($l,b$ =83.0\degr,--0.1\degr),
respectively. In the extraction only sources with quality flag `--1'
(stellar) and `--2' (probably stellar) have been taken into account
and the condition was set that the sources were detected in both the
direct as well as the offset fields.  In the colour-colour diagrams we
overplot the colour-tracks for unreddened data as well as for
$E(B-V)$=2.0 and 4.0.

In field 6160 (Fig.\ \ref{fig:uvex6160}, $l=79.6$\degr, $b=-2.8$\degr)
it can be clearly seen that the main-sequence stars are reddened
($E(B-V) = 1.25$ according to Schlegel, Finkbeiner\& Davis, 1998).  On
the blue side a small number of blue excess sources are present,
varying in magnitude between $18.5 < g <22.5$ and at $g-r \sim
0$. These are unreddened, intrinsically blue and intrinsically
low-luminosity objects that lie in front of the bulk of the main
sequence population.  These are the `UV-excess' sources that give
their name to the survey: predominantly white dwarfs and white dwarf
binaries. The reddening of the main-sequence causes the bulk of the
stars to shift to redder colours overall, uncovering a population of
`warm' ($T <$ 10\,000 K) white dwarfs. In unreddened (higher
galactic latitude) fields these `warm' white dwarfs merge with the
main-sequence and become difficult to identify in broad-band
photometry.  Due to the shallower depth of the $He${\sc i}
observations the faintest UV-excess sources in $g$ are not detected in
the $He${\sc i} filter (Fig.\ \ref{fig:uvex6160}).  Due to their blue
colour most are detected in the $U$-band. A distinction between DA and
DB white dwarfs can already be made on the basis of the $(U-g)$
vs. $(g-r)$ diagram, but will be further aided by the ($He${\sc
  i}$-r$) vs. $(g-r)$ and the (\heir) vs. ($r-H\alpha$) diagrams.

In field 6167 (Fig.\ \ref{fig:uvex6167}) the reddening is higher
($E(B-V)$=3.10 according to Schlegel, Finkbeiner \& Davis, 1998),
which is not surprising given its location in the mid-plane.  The
reddening is such that all stars earlier than M0 are substiantially
reddened.  In the (\heir) vs. $(g-r)$ diagram M-type stars show a
distinctive down-turn in the \heir colour, making them easily
identifiable. The same stars can be seen as the almost vertical
sequence at $g-r$=1.5 and running from $19 < g <
23$. Counter-intuitively the intrinsically faint late-type M-stars
have become some of the {\sl bluest} objects in the field, apart from
the real stellar remnants located bluewards of $g-r < $1 and $g > 19$.

\section{Seeing statistics \& Crowding}
\label{sec:seeing}

\begin{figure*}
\centerline{\epsfig{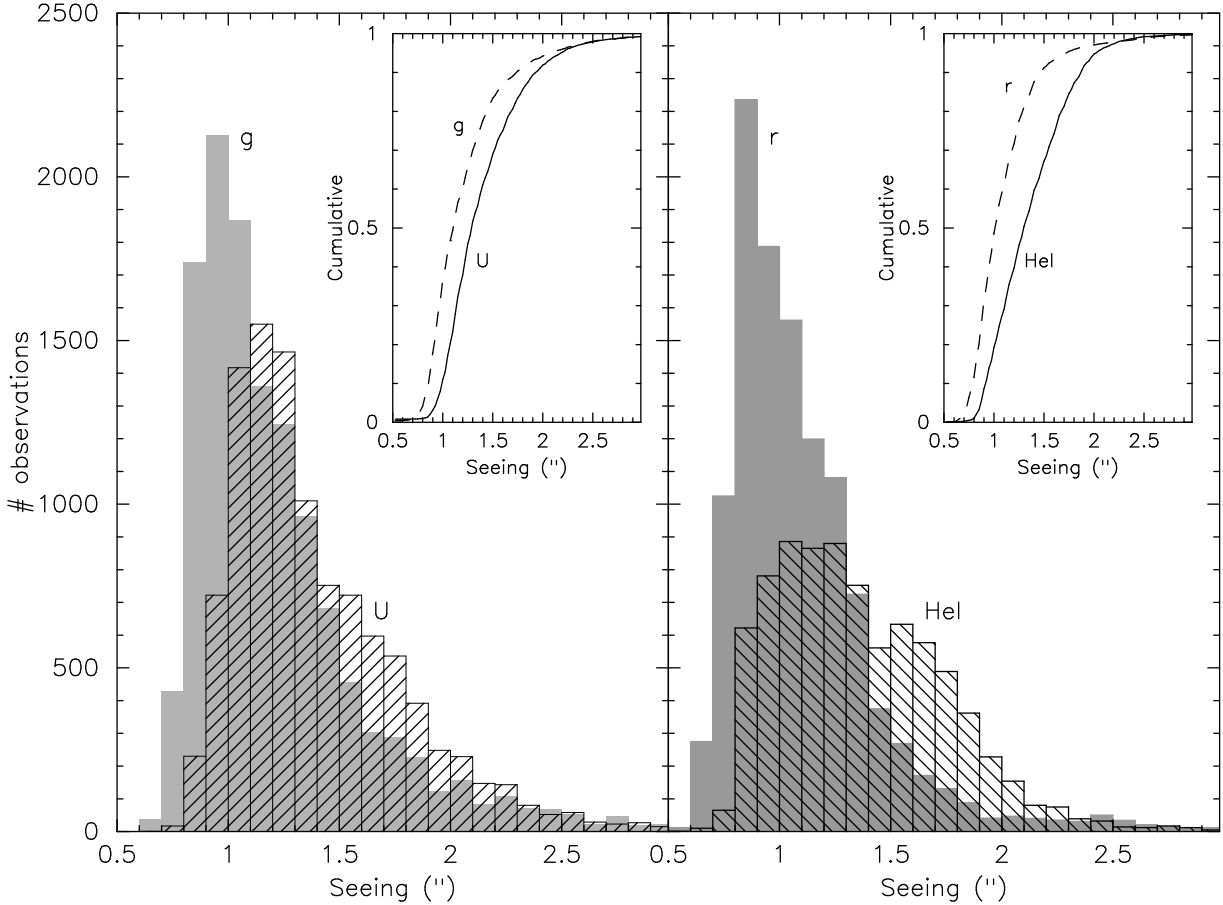}}
\caption[]{Seeing distributions for \UVEX observations in the period
  June 2006 - November 2007, for the $U$- and $g$-band observations
  (left panel) and the $r$- and $He${\sc i}-band observations (right
  panel). Inserts show the cumulative distributions. 
  \label{fig:seeing}}
\end{figure*}

For all data up to November 2007 we have collected the seeing
statistics in the four {\sl UVEX} filters
(Fig.\ \ref{fig:seeing}). This is for a total of $\sim$375 square
degrees and $\sim$3\,000 pointings over the period June 2006 -
November 2007. It can be seen from Fig.\ \ref{fig:seeing} that the
median seeing in the {\sl UVEX} data so far is
(1\farcs3,1\farcs1,1\farcs0,1\farcs4) for the ($U,g,r,He${\sc i})
filters, respectively. The $He${\sc i} data shows a qualitatively
different behaviour than the other three bands with a much broader
maximum. This is most likely caused by the fact that most of the
$He${\sc i} data is taken with an integration time of 180 seconds, but
with no autoguider. This causes small errors in the telescope
tracking, which translate into a deteriorated seeing. 

\begin{figure}
\centerline{\epsfig{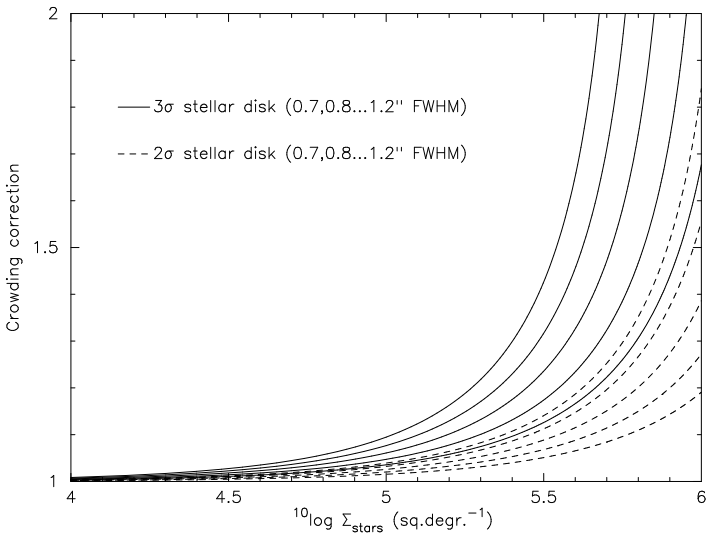}}
\caption[]{The crowding correction factor ($f^\prime/f$ in the Eq. 4
  of Irwin \& Trimble, 1984) between observed versus actual number
  densities of stars for a range in seeing fwhm's (0.7, 0.8...1.2:
  lines from bottom to top resp.) and for two settings of the stellar
  image threshold radius at 2$\sigma$ (dashed lines) and 3$\sigma$
  (full lines) Gaussian e-folding lengths.
  \label{fig:crowding}
}
\end{figure}

With the stellar densities expected and detected in the Galactic Plane
down to g$\sim$22 crowding becomes a real concern for number
statistics studies. Surface densities of detected point sources in the
{\sl UVEX} fields can reach up to 200\,000 sources per pointing,
i.e. 700\,000 stars per square degree. Following Irwin \& Trimble
(1984) we here make a global estimate of the effect of crowding in the
{\sl UVEX} data. Using their Eq.\,4 and inserting the relevant numbers
for {\sl UVEX} we calculate the crowding correction factor (i.e. their
$f^\prime/f$) for seeing disk FWHM values of 0\farcs7 - 1\farcs2
as shown in Fig.\ \ref{fig:crowding}. As can be seen the crowding
factor is a strong function of the actual seeing, and also of the
assumed radius of the actual stellar profile. Fig.\ \ref{fig:crowding}
shows the correction factors for both a 2$\sigma$ Gaussian profile
cut-off radius as well as a 3$\sigma$ cut-off. We see that at the
maximum density of detected sources in {\sl UVEX}, which is close to 1
million sources per square degree, we reach a crowding correction of
at least 20\% in the best of cases (0\farcs7 seeing, 2$\sigma$
cut-off) and quickly reach $>$100\% when a 3$\sigma$ cut-off radius is
taken. Of course in reality the actual crowding will also depend on
the actual magnitude difference between two nearby, almost overlapping
sources and will require a detailed field-to-field modeling, but this
global estimate shows that for the most crowded regions of the
Galactic Plane crowding is a serious issue.

\section{Conclusions}
\label{sec:conclusions}

The \UVEX survey offers the possibility to detect intrinsically blue,
faint objects in the Galactic Plane, as well as offers the first-ever
homogeneous blue survey of the Galactic Plane and is ideal for
uncovering a large population of stellar remnants. The depth of the
$g$-band observations, close to the ground-based confusion limit, will
allow for detailed Galactic structure research. The combination with
the \IPHAS survey offers the first ever optical survey of the Northern
Galactic Plane in the $U,g,r,i$, $H\alpha$ and \mbox{$He${\sc i}
  $5875$} filters. The Southern Plane will be covered by the {\sl
  VPHAS+} survey in the same bands (minus the $He${\sc i} band), on
the VLT Survey Telescope at the European Southern
Observatory. Combined, these three surveys form the heart of the
European Galactic Plane Surveys ({\sl EGAPS}). When completed {\sl
  EGAPS} will provide positions, colours for close to one billion
stars in our Galaxy. From the first study on proper motions from EGAPS
Deacon et al. (2009) showed that we can expect $\sim$140 objects per
square degree with a proper motion $\mu \geq$ 20 mas yr$^{-1}$.

\section*{Acknowledgement}

This paper makes use of data collected at the Isaac Newton Telescope,
operated on the island of La Palma by the Isaac Newton Group in the
Spanish Observatorio del Roque de los Muchachos of the Inst\'{\i}tuto
de Astrof\'{\i}sica de Canarias. We acknowledge the use of data
products from 2MASS, which is a joint project of the University of
Massachusetts and the Infrared Processing and Analysis
Center/California Institute of Technology (funded by the National
Aeronautics and Space Administration and National Science Foundation
of the USA). KV is supported by a NWO-EW grant 614.000.601 to PJG. ND
is supported by NOVA and NWO-VIDI grant 639.042.201 to PJG. The
authors would like to thank Detlev Koester and Pierre Bergeron for
making available their white dwarf models on which a significant part
of the colour simulations in this paper are based.

\label{lastpage}

\clearpage

\appendix

\oddsidemargin=-2cm
\evensidemargin=-2cm

\section{\UVEX colours for Main-sequence stars, including reddening}
\label{app:uvexcolours}

\begin{table*}
\caption[]{\UVEX colour indices $(U-g)$, $(g-r)$, $(HeI-r)$ for Pickles main-sequence stars including
  reddening. \label{tab:mainsequence} }
\small
\noindent
\begin{tabular}{p{3mm}@{\ \ \ }p{9mm}@{\ \ \ }p{9mm}@{\ \ \ }p{10mm}@{\ \ \ }p{9mm}@{\ \ \ }p{8mm}@{\ \ \ }p{10mm}@{\ \ \ }p{8mm}@{\ \ \ }p{8mm}@{\ \ \ }p{10mm}@{\ \ \ }p{8mm}@{\ \ \ }p{8mm}@{\ \ \ }p{10mm}@{\ \ \ }p{8mm}@{\ \ \ }p{8mm}@{\ \ \ }p{10mm}}
\hline 
\multicolumn{1}{l}{Spec.}&
\multicolumn{3}{l}{$E(B-V)=0.0$}&
\multicolumn{3}{l}{$E(B-V)=1.0$}&
\multicolumn{3}{l}{$E(B-V)=2.0$}&
\multicolumn{3}{l}{$E(B-V)=3.0$}&
\multicolumn{3}{l}{$E(B-V)=4.0$}
\\
type & $U$$-$$g$ & $g$$-$$r$ & $He${\sc i}$-$$r$ & $U$$-$$g$ & $g$$-$$r$ &
$He${\sc i}$-$$r$ & $U$$-$$g$ & $g$$-$$r$ & $He${\sc i}$-$$r$&
  $U$$-$$g$ & $g$$-$$r$ & $He${\sc i}$-$$r$& $U$$-$$g$ & $g$$-$$r$ &
  $He${\sc i}$-$$r$ \\ \hline
O5V & --1.434 & --0.294 &  0.006  &   --0.239 &  0.667 &  0.149  &   1.044 &  1.547 &  0.326&2.400 &  2.364 &  0.538 &   3.811 &  3.137 &  0.784     \\ 
O9V & --1.355 & --0.301 &  0.008  &   --0.153 &  0.652 &  0.151  &   1.134 &  1.525 &  0.329&2.492 &  2.338 &  0.541 &   3.903 &  3.111 &  0.788     \\
B0V & --1.293 & --0.265 &  0.006  &   --0.099 &  0.691 &  0.148  &   1.183 &  1.567 &  0.326&2.537 &  2.381 &  0.538 &   3.945 &  3.152 &  0.784     \\
B1V & --1.158 & --0.180 &  0.001  &    0.047 &  0.767 &  0.146   &   1.337 &  1.635 &  0.326&2.699 &  2.442 &  0.541 &   4.111 &  3.209 &  0.788    \\
B3V & --0.863 & --0.160 &  0.004  &    0.331 &  0.785 &  0.151   &   1.609 &  1.651 &  0.332&2.956 &  2.458 &  0.548 &   4.354 &  3.226 &  0.798    \\
B8V & --0.327 & --0.045 &  0.005  &    0.842 &  0.895 &  0.153   &   2.095 &  1.757 &  0.337&3.418 &  2.561 &  0.555 &   4.792 &  3.326 &  0.806    \\
B9V & --0.210 & --0.003 &  0.009  &    0.961 &  0.932 &  0.160   &   2.217 &  1.791 &  0.345&3.542 &  2.592 &  0.564 &   4.918 &  3.355 &  0.816    \\
A0V &  0.026 &  0.013 &  0.002  &    1.196 &  0.943 &  0.153  &   2.451 &  1.798 &  0.338   &3.775 &  2.596 &  0.558 &   5.149 &  3.357 &  0.811 \\
A2V &  0.085 &  0.039 & --0.002 &    1.253 &  0.969 &  0.149  &   2.504 &  1.823 &  0.336   &3.824 &  2.621 &  0.557 &   5.194 &  3.383 &  0.812  \\
A3V &  0.090 &  0.106 &  0.011  &    1.267 &  1.029 &  0.164  &   2.528 &  1.878 &  0.352   &3.856 &  2.672 &  0.575 &   5.233 &  3.431 &  0.831 \\
A5V &  0.148 &  0.150 &  0.017  &    1.329 &  1.066 &  0.171  &   2.594 &  1.909 &  0.361   &3.926 &  2.698 &  0.585 &   5.305 &  3.454 &  0.842 \\
A7V &  0.175 &  0.217 &  0.024  &    1.367 &  1.134 &  0.182  &   2.641 &  1.978 &  0.374   &3.981 &  2.769 &  0.600 &   5.368 &  3.526 &  0.859 \\
F0V &  0.158 &  0.325 &  0.046  &    1.370 &  1.235 &  0.208  &   2.662 &  2.074 &  0.405   &4.017 &  2.862 &  0.637 &   5.416 &  3.618 &  0.901 \\
F2V &  0.140 &  0.399 &  0.070  &    1.365 &  1.299 &  0.235  &   2.669 &  2.131 &  0.435   &4.035 &  2.914 &  0.669 &   5.444 &  3.667 &  0.935 \\
F5V &  0.130 &  0.460 &  0.068  &    1.369 &  1.355 &  0.234  &   2.685 &  2.182 &  0.436   &4.062 &  2.962 &  0.670 &   5.482 &  3.713 &  0.937 \\
F6V &  0.177 &  0.486 &  0.074  &    1.416 &  1.382 &  0.242  &   2.732 &  2.212 &  0.445   &4.109 &  2.994 &  0.682 &   5.529 &  3.747 &  0.951 \\
F8V &  0.269 &  0.552 &  0.086  &    1.514 &  1.443 &  0.256  &   2.836 &  2.268 &  0.462   &4.217 &  3.047 &  0.701 &   5.640 &  3.798 &  0.971 \\
G0V &  0.355 &  0.567 &  0.096  &    1.602 &  1.455 &  0.268  &   2.925 &  2.280 &  0.474   &4.307 &  3.060 &  0.714 &   5.731 &  3.811 &  0.985 \\
G2V &  0.438 &  0.631 &  0.100  &    1.695 &  1.514 &  0.273  &   3.028 &  2.335 &  0.481   &4.417 &  3.112 &  0.722 &   5.847 &  3.862 &  0.995 \\
G5V &  0.569 &  0.658 &  0.106  &    1.829 &  1.536 &  0.280  &   3.162 &  2.353 &  0.489   &4.552 &  3.127 &  0.732 &   5.983 &  3.875 &  1.005 \\
G8V &  0.695 &  0.734 &  0.130  &    1.972 &  1.610 &  0.307  &   3.319 &  2.427 &  0.519   &4.721 &  3.202 &  0.764 &   6.162 &  3.951 &  1.040 \\
K0V &  0.769 &  0.779 &  0.150  &    2.036 &  1.656 &  0.331  &   3.376 &  2.474 &  0.545   &4.771 &  3.250 &  0.793 &   6.204 &  4.000 &  1.071 \\
K2V &  1.072 &  0.913 &  0.164  &    2.360 &  1.780 &  0.346  &   3.717 &  2.591 &  0.563   &5.126 &  3.361 &  0.813 &   6.572 &  4.107 &  1.093 \\
K3V &  1.225 &  0.988 &  0.201  &    2.498 &  1.864 &  0.387  &   3.844 &  2.681 &  0.607   &5.246 &  3.454 &  0.859 &   6.688 &  4.201 &  1.143 \\
K4V &  1.356 &  1.113 &  0.235  &    2.648 &  1.976 &  0.425  &   4.009 &  2.784 &  0.649   &5.424 &  3.553 &  0.906 &   6.877 &  4.297 &  1.193 \\
K5V &  1.686 &  1.277 &  0.252  &    2.983 &  2.133 &  0.443  &   4.346 &  2.935 &  0.667   &5.760 &  3.698 &  0.923 &   7.210 &  4.436 &  1.209 \\
K7V &  1.721 &  1.411 &  0.328  &    3.032 &  2.250 &  0.526  &   4.407 &  3.038 &  0.758   &5.831 &  3.791 &  1.021 &   7.288 &  4.522 &  1.313 \\
M0V &  1.806 &  1.394 &  0.378  &    3.109 &  2.245 &  0.580  &   4.478 &  3.044 &  0.816   &5.895 &  3.806 &  1.083 &   7.347 &  4.545 &  1.379 \\
M1V &  1.872 &  1.389 &  0.425  &    3.174 &  2.227 &  0.628  &   4.541 &  3.015 &  0.865   &5.956 &  3.768 &  1.133 &   7.404 &  4.501 &  1.430 \\
M2V &  1.745 &  1.485 &  0.491  &    3.059 &  2.320 &  0.702  &   4.436 &  3.105 &  0.946   &5.859 &  3.857 &  1.220 &   7.313 &  4.590 &  1.523 \\
M3V &  1.821 &  1.496 &  0.626  &    3.143 &  2.320 &  0.848  &   4.525 &  3.098 &  1.102   &5.949 &  3.848 &  1.386 &   7.401 &  4.581 &  1.697 \\
M4V &  2.285 &  1.513 &  0.742  &    3.622 &  2.336 &  0.977  &   5.013 &  3.119 &  1.244   &6.442 &  3.876 &  1.542 &   7.895 &  4.619 &  1.866 \\
M5V &  2.230 &  1.605 &  0.769  &    3.583 &  2.423 &  1.006  &   4.986 &  3.202 &  1.275   &6.423 &  3.956 &  1.573 &   7.883 &  4.696 &  1.896 \\
M6V &  2.189 &  1.717 &  0.784  &    3.568 &  2.526 &  1.038  &   4.988 &  3.302 &  1.323   &6.437 &  4.059 &  1.635 &   7.904 &  4.803 &  1.972  \\ \hline   
\end{tabular}                                                                                
\end{table*}

\newpage

\begin{table*}
\label{tab:giants}
\caption[]{\UVEX colour indices $(U-g)$, $(g-r)$, $(HeI-r)$ for Pickles Giants including reddening.}
\small
\begin{tabular}{p{10mm}@{\ \ \ }p{9mm}@{\ \ \ }p{9mm}@{\ \ \ }p{10mm}@{\ \ \ }p{9mm}@{\ \ \ }p{7mm}@{\ \ \ }p{10mm}@{\ \ \ }p{7mm}@{\ \ \ }p{7mm}@{\ \ \ }p{10mm}@{\ \ \ }p{7mm}@{\ \ \ }p{7mm}@{\ \ \ }p{10mm}@{\ \ \ }p{7mm}@{\ \ \ }p{7mm}@{\ \ \ }p{10mm}}
\hline 
\multicolumn{1}{l}{Spec.}&
\multicolumn{3}{l}{$E(B-V)$=0.0}&
\multicolumn{3}{l}{$E(B-V)$=1.0}&
\multicolumn{3}{l}{$E(B-V)$=2.0}&
\multicolumn{3}{l}{$E(B-V)$=3.0}&
\multicolumn{3}{l}{$E(B-V)$=4.0}\\
type    & $U$$-$$g$ & $g$$-$$r$ & $He${\sc i}$-$$r$ &$U$$-$$g$ &
$g$$-$$r$ & $He${\sc i}$-$$r$ & $U$$-$$g$ & $g$$-$$r$ & $He${\sc
  i}$-$$r$& $U$$-$$g$ & $g$$-$$r$ & $He${\sc i}$-$$r$& $U$$-$$g$ &
$g$$-$$r$ & $He${\sc i}$-$$r$  \\ \hline
O8III   & --1.382 & --0.259 &  0.001  &  -0.181 &  0.696 &  0.143  &   1.107 &  1.571 &  0.320&    2.468 &  2.383 &  0.532  &   3.881 &  3.154 &  0.777  \\
B1-2III & --1.170 & --0.215 &  0.016  &   0.029 &  0.733 &  0.160  &   1.315 &  1.603 &  0.340&    2.671 &  2.412 &  0.554  &   4.078 &  3.181 &  0.801  \\
B3III   & --0.852 & --0.185 &  0.032  &   0.344 &  0.759 &  0.179  &   1.626 &  1.627 &  0.361&   2.976 &  2.435 &  0.578 &   4.378 &  3.204 &  0.828     \\
B5III   & --0.619 & --0.132 &  0.010  &   0.568 &  0.812 &  0.158  &   1.838 &  1.679 &  0.341&  3.178 &  2.487 &  0.559  &   4.568 &  3.256 &  0.810     \\
B9III   & --0.148 & --0.045 & -0.002  &   1.012 &  0.895 &  0.147  &   2.256 &  1.757 &  0.330&  3.571 &  2.561 &  0.548  &   4.938 &  3.326 &  0.799     \\
A0III   &  0.013 &  0.040 & --0.003  &   1.188 &  0.968 &  0.147  &   2.447 &  1.821 &  0.333 &  3.774 &  2.618 &  0.552  &   5.151 &  3.377 &  0.805     \\
A3III   &  0.173 &  0.136 & --0.008  &   1.348 &  1.054 &  0.145  &   2.605 &  1.899 &  0.334 &  3.930 &  2.689 &  0.556  &   5.304 &  3.445 &  0.812     \\
A5III   &  0.185 &  0.172 &  0.031  &   1.370 &  1.088 &  0.186  &   2.638 &  1.932 &  0.377  &  3.972 &  2.722 &  0.603  &   5.353 &  3.480 &  0.861   \\
A7III   &  0.230 &  0.234 &  0.025  &   1.418 &  1.150 &  0.183  &   2.688 &  1.994 &  0.376  &  4.024 &  2.784 &  0.602  &   5.408 &  3.541 &  0.862   \\
F0III   &  0.241 &  0.276 &  0.038  &   1.435 &  1.186 &  0.198  &   2.711 &  2.026 &  0.393  &  4.051 &  2.815 &  0.622  &   5.439 &  3.571 &  0.883   \\
F2III   &  0.198 &  0.437 &  0.060  &   1.417 &  1.337 &  0.226  &   2.717 &  2.169 &  0.427  &  4.079 &  2.952 &  0.662  &   5.486 &  3.704 &  0.929   \\
F5III   &  0.233 &  0.451 &  0.074  &   1.456 &  1.352 &  0.242  &   2.758 &  2.186 &  0.444  &  4.122 &  2.971 &  0.680  &   5.530 &  3.726 &  0.948   \\
G0III   &  0.555 &  0.663 &  0.106  &   1.812 &  1.544 &  0.281  &   3.144 &  2.364 &  0.490  &  4.533 &  3.140 &  0.732  &   5.961 &  3.888 &  1.006   \\
G5III   &  0.969 &  0.822 &  0.127  &   2.256 &  1.686 &  0.307  &   3.612 &  2.493 &  0.522  &  5.019 &  3.261 &  0.770  &   6.462 &  4.007 &  1.048   \\
G8III   &  1.134 &  0.883 &  0.133  &   2.419 &  1.740 &  0.314  &   3.772 &  2.543 &  0.529  &  5.176 &  3.307 &  0.777  &   6.616 &  4.049 &  1.055   \\
K0III   &  1.313 &  0.891 &  0.162  &   2.597 &  1.753 &  0.346  &   3.948 &  2.559 &  0.564  &  5.351 &  3.328 &  0.815  &   6.789 &  4.074 &  1.096   \\
K1III   &  1.462 &  0.972 &  0.162  &   2.754 &  1.826 &  0.346  &   4.111 &  2.628 &  0.565  &  5.517 &  3.393 &  0.816  &   6.958 &  4.137 &  1.098   \\
K2III   &  1.600 &  1.047 &  0.186  &   2.885 &  1.897 &  0.374  &   4.235 &  2.696 &  0.597  &  5.635 &  3.458 &  0.852  &   7.071 &  4.200 &  1.137   \\
K3III   &  1.852 &  1.118 &  0.191  &   3.148 &  1.960 &  0.380  &   4.507 &  2.753 &  0.603  &  5.915 &  3.513 &  0.859  &   7.356 &  4.252 &  1.145   \\
K4III   &  2.272 &  1.303 &  0.198  &   3.601 &  2.123 &  0.391  &   4.989 &  2.900 &  0.619  &  6.419 &  3.647 &  0.878  &   7.878 &  4.377 &  1.168   \\
K5III   &  2.388 &  1.345 &  0.252  &   3.700 &  2.175 &  0.452  &   5.072 &  2.959 &  0.684  &  6.489 &  3.712 &  0.949  &   7.936 &  4.448 &  1.244   \\
M0III   &  2.651 &  1.444 &  0.327  &   3.969 &  2.266 &  0.532  &   5.345 &  3.044 &  0.770  &  6.763 &  3.793 &  1.041  &   8.211 &  4.527 &  1.341   \\
M1III   &  2.585 &  1.449 &  0.313  &   3.893 &  2.272 &  0.518  &   5.263 &  3.050 &  0.757  &  6.678 &  3.800 &  1.028  &   8.125 &  4.533 &  1.328   \\
M2III   &  2.710 &  1.513 &  0.385  &   4.030 &  2.337 &  0.597  &   5.408 &  3.117 &  0.844  &  6.828 &  3.870 &  1.122  &   8.276 &  4.607 &  1.429   \\
M3III   &  2.608 &  1.477 &  0.470  &   3.923 &  2.302 &  0.688  &   5.298 &  3.084 &  0.939  &  6.716 &  3.838 &  1.221  &   8.163 &  4.578 &  1.532   \\
M4III   &  2.557 &  1.507 &  0.629  &   3.869 &  2.335 &  0.858  &   5.242 &  3.120 &  1.121  &    6.656 &  3.880 &  1.415 &   8.099 &  4.625 &  1.736 \\
M5III   &  1.975 &  1.547 &  0.757  &   3.272 &  2.390 &  0.997  &   4.636 &  3.186 &  1.270  &    6.047 &  3.954 &  1.573 &   7.488 &  4.707 &  1.903 \\ \hline    
\end{tabular} 
\end{table*}



\newpage

\begin{table*}

\caption[]{\UVEX/\IPHAS colour indices $(U-g)$, $(g-r)$, $(HeI-r)$ for Pickles Supergiants including reddening.}

\begin{tabular}{p{3mm}@{\ \ \ }p{9mm}@{\ \ \ }p{9mm}@{\ \ \ }p{10mm}@{\ \ \ }p{9mm}@{\ \ \ }p{9mm}@{\ \ \ }p{10mm}@{\ \ \ }p{8mm}@{\ \ \ }p{8mm}@{\ \ \ }p{10mm}@{\ \ \ }p{8mm}@{\ \ \ }p{8mm}@{\ \ \ }p{10mm}@{\ \ \ }p{8mm}@{\ \ \ }p{8mm}@{\ \ \ }p{10mm}}
\hline 
\multicolumn{1}{l}{Spec.}&
\multicolumn{3}{l}{$E(B-V)$=0.0}&
\multicolumn{3}{l}{$E(B-V)$=1.0}&
\multicolumn{3}{l}{$E(B-V)$=2.0}&
\multicolumn{3}{l}{$E(B-V)$=3.0}&
\multicolumn{3}{l}{$E(B-V)$=4.0}\\
type & $U$$-$$g$ & $g$$-$$r$ & $He${\sc i}$-$$r$ &$U$$-$$g$ &
$g$$-$$r$ & $He${\sc i}$-$$r$ & $U$$-$$g$ & $g$$-$$r$ &$He${\sc
  i}$-$$r$ & $U$$-$$g$ & $g$$-$$r$ & $He${\sc i}$-$$r$ & $U$$-$$g$ &
$g$$-$$r$ & $He${\sc i}$-$$r$\\ \hline
B0I & --1.250 & --0.193 &  0.040  &  -0.034 &  0.748 &  0.186 &   1.267 &1.613 & 0.368&  2.635 &  2.420 &  0.585 &   4.053 &  3.189 &  0.834    \\
B1I & --1.159 & --0.118 &  0.034  &   0.056 &  0.825 &  0.186 &   1.354 &1.692 & 0.374&  2.720 &  2.502 &  0.596 &   4.134 &  3.274 &  0.851    \\
B3I & --0.951 & --0.064 &  0.041  &   0.251 &  0.878 &  0.190 &   1.539 &1.741 & 0.374&  2.896 &  2.545 &  0.593 &   4.303 &  3.309 &  0.845    \\
B5I & --0.852 & --0.019 &  0.052 &   0.358 &  0.913 &  0.206 &   1.651 &1.771 & 0.395 &  3.010 &  2.573 &  0.619 &   4.417 &  3.339 &  0.875   \\
B8I & --0.659 &  0.005 &  0.067 &   0.543 &  0.931 &  0.221  &   1.828 &1.783 & 0.410 &  3.179 &  2.581 &  0.634 &   4.578 &  3.344 &  0.890  \\
A0I & --0.264 &  0.047 &  0.051 &   0.902 &  0.977 &  0.206  &   2.153 &1.832 &  0.397&  3.474 &  2.632 &  0.621 &   4.845 &  3.397 &  0.879   \\
A2I & --0.217 &  0.143 &  0.046 &   0.964 &  1.064 &  0.202  &   2.229 &1.912 &  0.394&  3.563 &  2.706 &  0.620 &   4.946 &  3.465 &  0.878   \\
F0I &  0.373 &  0.224 &  0.048 &   1.526 &  1.141 &  0.208 &   2.764 &1.986 &  0.403  &  4.071 &  2.780 &  0.633 &   5.429 &  3.541 &  0.896  \\ 
F5I &  0.378 &  0.288 &  0.069 &   1.543 &  1.204 &  0.234 &   2.793 &2.050 &  0.433  &  4.111 &  2.846 &  0.666 &   5.480 &  3.609 &  0.931  \\ 
F8I &  0.728 &  0.558 &  0.077 &   1.938 &  1.438 &  0.244 &   3.225 &2.255 &  0.446  &  4.574 &  3.029 &  0.681 &   5.967 &  3.776 &  0.948  \\ 
G0I &  0.802 &  0.740 &  0.101 &   2.052 &  1.610 &  0.274 &   3.377 &2.420 &  0.482  &  4.757 &  3.190 &  0.724 &   6.178 &  3.934 &  0.997  \\ 
G2I &  1.052 &  0.825 &  0.090 &   2.320 &  1.682 &  0.265 &   3.658 &2.483 &  0.475  &  5.049 &  3.246 &  0.718 &   6.477 &  3.986 &  0.993  \\ 
G5I &  1.330 &  0.945 &  0.114 &   2.617 &  1.785 &  0.293 &   3.969 &2.575 &  0.507  &  5.369 &  3.331 &  0.754 &   6.804 &  4.067 &  1.032  \\ 
G8I &  1.773 &  1.108 &  0.183 &   3.062 &  1.948 &  0.371 &   4.416 &2.738 &  0.593  &  5.818 &  3.496 &  0.848 &   7.254 &  4.234 &  1.133  \\ 
K2I &  2.377 &  1.342 &  0.191 &   3.703 &  2.153 &  0.384 &   5.085 &2.922 &  0.611  &  6.508 &  3.665 &  0.870 &   7.960 &  4.392 &  1.159  \\ 
K3I &  2.473 &  1.442 &  0.243 &   3.799 &  2.255 &  0.442 &   5.183 &3.025 &  0.675  &  6.608 &  3.769 &  0.939 &   8.064 &  4.498 &  1.234  \\ 
K4I &  2.503 &  1.524 &  0.276 &   3.851 &  2.329 &  0.479 &   5.252 &3.095 &  0.715  &  6.692 &  3.837 &  0.983 &   8.158 &  4.565 &  1.281  \\ 
M2I &  2.676 &  1.624 &  0.483 &   4.002 &  2.433 &  0.703 &   5.383 &3.205 &  0.956  &  6.804 &  3.954 &  1.240 &   8.254 &  4.690 &  1.553  \\  \hline    
\end{tabular} 
\end{table*}


\newpage

\begin{table*}
\caption[]{\UVEX/\IPHAS colour indices $(U-g)$, $(g-r)$, $(He${\sc i}$-r)$ $(r-H\alpha)$ and $(r-i)$ for log(g)=8.0 Bergeron DA white dwarfs including reddening.}
\begin{tabular}{p{3mm}p{9mm}p{9mm}p{9mm}p{9mm}p{9mm}p{7mm}p{9mm}p{9mm}p{9mm}p{9mm}p{7mm}p{7mm}p{9mm}p{7mm}p{9mm}}
\hline 
\multicolumn{1}{l}{T (K)}&
\multicolumn{5}{l}{E(B--V)=0.0}&
\multicolumn{5}{l}{E(B--V)=1.0}&
\multicolumn{5}{l}{E(B--V)=2.0}\\
   & $U$$-$$g$ & $g$$-$$r$ & $He${\sc i}$-$$r$ & $r$$-$$H\alpha$ & $r$$-$$i$ & $U$$-$$g$ & $g$$-$$r$ &
$He${\sc i}$-$$r$ & $r$$-$$H\alpha$ & $r$$-$$i$ &$U$$-$$g$ & $g$$-$$r$
& $He${\sc i}$-$$r$ & $r$$-$$H\alpha$
& $r$$-$$i$ \\ \hline
 1500 &  1.809 &  0.761 & --0.232 & --0.275 & --2.272  &    3.113 &  1.469 & --0.132 & --0.030 & --1.598  &   4.475 &  2.129 & --0.004 &  0.186 & --0.901 \\
 1750 &  1.592 &  1.025 & --0.095 &  0.007 & --1.613  &    2.895 &  1.772 &  0.036 &  0.220 & --0.944  &   4.258 &  2.469 &  0.196 &  0.404 & --0.254 \\
 2000 &  1.419 &  1.160 &  0.029 &  0.215 & --1.036  &    2.719 &  1.939 &  0.185 &  0.403 & --0.375  &   4.082 &  2.669 &  0.372 &  0.561 &  0.303 \\
 2250 &  1.271 &  1.229 &  0.118 &  0.325 & --0.531  &    2.570 &  2.034 &  0.293 &  0.495 &  0.124  &   3.931 &  2.788 &  0.499 &  0.634 &  0.792 \\
 2500 &  1.141 &  1.254 &  0.174 &  0.375 & --0.093  &    2.437 &  2.078 &  0.361 &  0.534 &  0.562  &   3.798 &  2.849 &  0.580 &  0.660 &  1.227 \\
 2750 &  1.030 &  1.249 &  0.203 &  0.393 &  0.258  &    2.324 &  2.084 &  0.396 &  0.545 &  0.920  &   3.684 &  2.867 &  0.622 &  0.664 &  1.590 \\
 3000 &  0.936 &  1.225 &  0.214 &  0.396 &  0.501  &    2.228 &  2.068 &  0.409 &  0.546 &  1.174  &   3.586 &  2.858 &  0.638 &  0.662 &  1.853 \\
 3250 &  0.850 &  1.191 &  0.214 &  0.393 &  0.638  &    2.140 &  2.040 &  0.410 &  0.542 &  1.320  &   3.497 &  2.834 &  0.639 &  0.658 &  2.007 \\
 3500 &  0.768 &  1.151 &  0.209 &  0.386 &  0.694  &    2.056 &  2.003 &  0.404 &  0.536 &  1.381  &   3.412 &  2.800 &  0.633 &  0.653 &  2.073 \\
 3750 &  0.687 &  1.106 &  0.201 &  0.378 &  0.700  &    1.972 &  1.962 &  0.395 &  0.529 &  1.390  &   3.326 &  2.762 &  0.622 &  0.647 &  2.085 \\
 4000 &  0.603 &  1.057 &  0.191 &  0.368 &  0.682  &    1.885 &  1.916 &  0.383 &  0.522 &  1.373  &   3.238 &  2.718 &  0.609 &  0.641 &  2.070 \\
 4250 &  0.512 &  1.002 &  0.180 &  0.357 &  0.650  &    1.791 &  1.865 &  0.370 &  0.513 &  1.342  &   3.141 &  2.670 &  0.594 &  0.634 &  2.039 \\
 4500 &  0.412 &  0.941 &  0.168 &  0.344 &  0.610  &    1.688 &  1.808 &  0.355 &  0.502 &  1.303  &   3.036 &  2.615 &  0.577 &  0.626 &  2.001 \\
 4750 &  0.307 &  0.876 &  0.155 &  0.329 &  0.567  &    1.580 &  1.747 &  0.340 &  0.490 &  1.261  &   2.925 &  2.558 &  0.559 &  0.616 &  1.959 \\
 5000 &  0.201 &  0.810 &  0.143 &  0.311 &  0.523  &    1.470 &  1.686 &  0.324 &  0.474 &  1.218  &   2.812 &  2.499 &  0.541 &  0.603 &  1.917 \\
 5250 &  0.100 &  0.747 &  0.130 &  0.288 &  0.482  &    1.365 &  1.626 &  0.309 &  0.453 &  1.178  &   2.705 &  2.443 &  0.523 &  0.585 &  1.877 \\
 5500 &  0.006 &  0.688 &  0.118 &  0.260 &  0.445  &    1.268 &  1.572 &  0.295 &  0.428 &  1.141  &   2.606 &  2.391 &  0.506 &  0.562 &  1.842 \\
 6000 & --0.160 &  0.584 &  0.098 &  0.205 &  0.381  &    1.097 &  1.474 &  0.270 &  0.377 &  1.079  &   2.430 &  2.298 &  0.477 &  0.516 &  1.780 \\
 6500 & --0.303 &  0.494 &  0.080 &  0.157 &  0.327  &    0.949 &  1.390 &  0.249 &  0.334 &  1.026  &   2.279 &  2.219 &  0.452 &  0.476 &  1.728 \\
 7000 & --0.409 &  0.420 &  0.066 &  0.117 &  0.280  &    0.839 &  1.321 &  0.232 &  0.297 &  0.980  &   2.167 &  2.152 &  0.432 &  0.442 &  1.684 \\
 7500 & --0.479 &  0.360 &  0.055 &  0.082 &  0.240  &    0.766 &  1.264 &  0.217 &  0.265 &  0.941  &   2.092 &  2.098 &  0.415 &  0.412 &  1.646 \\
 8000 & --0.520 &  0.310 &  0.044 &  0.048 &  0.204  &    0.723 &  1.216 &  0.204 &  0.233 &  0.906  &   2.047 &  2.052 &  0.400 &  0.383 &  1.611 \\
 8500 & --0.539 &  0.266 &  0.034 &  0.014 &  0.171  &    0.702 &  1.174 &  0.192 &  0.202 &  0.873  &   2.025 &  2.010 &  0.385 &  0.354 &  1.579 \\
 9000 & --0.543 &  0.228 &  0.023 & --0.022 &  0.140  &    0.698 &  1.136 &  0.178 &  0.167 &  0.843  &   2.020 &  1.972 &  0.369 &  0.322 &  1.550 \\
 9500 & --0.537 &  0.193 &  0.011 & --0.068 &  0.110  &    0.703 &  1.100 &  0.164 &  0.124 &  0.814  &   2.025 &  1.935 &  0.352 &  0.282 &  1.522 \\
10000 & --0.528 &  0.160 & --0.000 & --0.119 &  0.083  &    0.713 &  1.067 &  0.150 &  0.076 &  0.788  &   2.035 &  1.901 &  0.335 &  0.236 &  1.497\\
10500 & --0.521 &  0.132 & --0.011 & --0.161 &  0.057  &    0.719 &  1.037 &  0.137 &  0.037 &  0.764  &   2.041 &  1.869 &  0.320 &  0.199 &  1.474\\
11000 & --0.522 &  0.106 & --0.019 & --0.191 &  0.035  &    0.718 &  1.010 &  0.127 &  0.008 &  0.742  &   2.040 &  1.842 &  0.308 &  0.173 &  1.453\\
11500 & --0.529 &  0.082 & --0.026 & --0.212 &  0.016  &    0.711 &  0.987 &  0.119 & --0.011 &  0.724  &   2.033 &  1.818 &  0.298 &  0.155 &  1.435\\
12000 & --0.536 &  0.061 & --0.031 & --0.224 & --0.001  &    0.704 &  0.966 &  0.112 & --0.021 &  0.708  &   2.026 &  1.797 &  0.290 &  0.146 &  1.420\\
12500 & --0.543 &  0.044 & --0.034 & --0.225 & --0.013  &    0.696 &  0.949 &  0.108 & --0.022 &  0.696  &   2.017 &  1.781 &  0.286 &  0.146 &  1.408\\
13000 & --0.554 &  0.027 & --0.036 & --0.223 & --0.024  &    0.684 &  0.933 &  0.105 & --0.020 &  0.685  &   2.004 &  1.766 &  0.282 &  0.149 &  1.398\\
13500 & --0.572 &  0.010 & --0.038 & --0.221 & --0.034  &    0.664 &  0.918 &  0.103 & --0.017 &  0.675  &   1.983 &  1.752 &  0.279 &  0.153 &  1.387\\
14000 & --0.597 & --0.005 & --0.039 & --0.217 & --0.044  &    0.639 &  0.905 &  0.102 & --0.013 &  0.665  &   1.957 &  1.740 &  0.278 &  0.157 &  1.378\\
14500 & --0.624 & --0.019 & --0.039 & --0.212 & --0.051  &    0.609 &  0.893 &  0.102 & --0.007 &  0.657  &   1.926 &  1.730 &  0.277 &  0.163 &  1.370\\
15000 & --0.653 & --0.031 & --0.039 & --0.206 & --0.058  &    0.579 &  0.882 &  0.102 & --0.001 &  0.650  &   1.895 &  1.721 &  0.277 &  0.169 &  1.363\\
15500 & --0.683 & --0.042 & --0.038 & --0.199 & --0.064  &    0.548 &  0.873 &  0.102 &  0.006 &  0.644  &   1.863 &  1.713 &  0.278 &  0.176 &  1.356\\
16000 & --0.712 & --0.052 & --0.038 & --0.192 & --0.070  &    0.518 &  0.865 &  0.102 &  0.013 &  0.638  &   1.832 &  1.706 &  0.278 &  0.183 &  1.350\\
16500 & --0.741 & --0.061 & --0.038 & --0.184 & --0.075  &    0.488 &  0.857 &  0.103 &  0.020 &  0.633  &   1.801 &  1.700 &  0.278 &  0.190 &  1.345\\
17000 & --0.769 & --0.070 & --0.038 & --0.177 & --0.079  &    0.460 &  0.850 &  0.103 &  0.027 &  0.629  &   1.772 &  1.695 &  0.279 &  0.197&  1.340  \\ \hline  
\end{tabular} 
\end{table*}

\newpage

\addtocounter{table}{-1}
\begin{table*}
\caption[]{, continued}
\flushleft
\begin{tabular}{ccccccccccc}
\hline 
\multicolumn{1}{l}{T (K)}&
\multicolumn{5}{l}{E(B--V)=3.0}&
\multicolumn{5}{l}{E(B--V)=4.0}\\
      & $U$$-$$g$ & $g$$-$$r$ & $He${\sc i}$-$$r$ & $r$$-$$H\alpha$ &
$r$$-$$i$ & $U$$-$$g$ & $g$$-$$r$ & $He${\sc i}$-$$r$ & $r$$-$$H\alpha$ & $r$$-$$i$\\ \hline
 1500 &  5.879 &  2.761 &  0.153 &  0.374 & -0.179  &   7.312 &  3.380 &  0.339 &  0.532 &  0.566 \\        
 1750 &  5.665 &  3.136 &  0.385 &  0.559 &  0.455  &   7.101 &  3.789 &  0.604 &  0.685 &  1.184 \\        
 2000 &  5.489 &  3.368 &  0.589 &  0.689 &  0.998  &   6.926 &  4.050 &  0.834 &  0.788 &  1.708 \\        
 2250 &  5.338 &  3.511 &  0.736 &  0.742 &  1.474  &   6.776 &  4.216 &  1.002 &  0.821 &  2.169 \\        
 2500 &  5.205 &  3.589 &  0.830 &  0.755 &  1.903  &   6.642 &  4.310 &  1.110 &  0.821 &  2.590 \\        
 2750 &  5.090 &  3.617 &  0.880 &  0.752 &  2.269  &   6.528 &  4.348 &  1.166 &  0.811 &  2.955 \\        
 3000 &  4.992 &  3.614 &  0.899 &  0.747 &  2.538  &   6.430 &  4.350 &  1.189 &  0.802 &  3.229 \\        
 3250 &  4.903 &  3.593 &  0.901 &  0.742 &  2.699  &   6.341 &  4.332 &  1.191 &  0.797 &  3.396 \\        
 3500 &  4.816 &  3.562 &  0.893 &  0.737 &  2.770  &   6.254 &  4.303 &  1.184 &  0.793 &  3.471 \\        
 3750 &  4.730 &  3.525 &  0.881 &  0.733 &  2.784  &   6.168 &  4.267 &  1.171 &  0.789 &  3.488 \\        
 4000 &  4.641 &  3.483 &  0.867 &  0.729 &  2.770  &   6.078 &  4.226 &  1.154 &  0.786 &  3.475 \\        
 4250 &  4.543 &  3.436 &  0.850 &  0.723 &  2.741  &   5.980 &  4.180 &  1.136 &  0.783 &  3.446 \\        
 4500 &  4.436 &  3.383 &  0.830 &  0.717 &  2.703  &   5.872 &  4.128 &  1.114 &  0.779 &  3.409 \\        
 4750 &  4.323 &  3.327 &  0.810 &  0.710 &  2.661  &   5.759 &  4.073 &  1.092 &  0.773 &  3.368 \\        
 5000 &  4.209 &  3.271 &  0.790 &  0.699 &  2.620  &   5.644 &  4.017 &  1.069 &  0.765 &  3.327 \\        
 5250 &  4.100 &  3.216 &  0.770 &  0.684 &  2.581  &   5.534 &  3.964 &  1.047 &  0.752 &  3.289 \\        
 5500 &  4.000 &  3.166 &  0.751 &  0.663 &  2.546  &   5.433 &  3.914 &  1.026 &  0.733 &  3.254 \\        
 6000 &  3.822 &  3.076 &  0.718 &  0.621 &  2.486  &   5.254 &  3.826 &  0.989 &  0.694 &  3.195 \\        
 6500 &  3.669 &  2.999 &  0.689 &  0.584 &  2.435  &   5.101 &  3.750 &  0.958 &  0.661 &  3.145 \\        
 7000 &  3.555 &  2.935 &  0.666 &  0.553 &  2.392  &   4.986 &  3.687 &  0.932 &  0.633 &  3.103 \\        
 7500 &  3.479 &  2.882 &  0.647 &  0.526 &  2.354  &   4.910 &  3.635 &  0.911 &  0.608 &  3.066 \\        
 8000 &  3.433 &  2.837 &  0.629 &  0.499 &  2.320  &   4.864 &  3.589 &  0.891 &  0.583 &  3.032 \\        
 8500 &  3.411 &  2.795 &  0.612 &  0.473 &  2.289  &   4.841 &  3.548 &  0.872 &  0.559 &  3.002 \\        
 9000 &  3.405 &  2.756 &  0.594 &  0.443 &  2.260  &   4.836 &  3.507 &  0.851 &  0.531 &  2.974 \\        
 9500 &  3.411 &  2.718 &  0.574 &  0.405 &  2.234  &   4.842 &  3.468 &  0.830 &  0.495 &  2.948 \\        
10000 &   3.421 &  2.682 &  0.555 &  0.362 &  2.209  &   4.852 &  3.430 &  0.809 &  0.455 &  2.925 \\       
10500 &   3.428 &  2.649 &  0.538 &  0.327 &  2.187  &   4.859 &  3.396 &  0.789 &  0.422 &  2.903 \\       
11000 &   3.427 &  2.621 &  0.524 &  0.302 &  2.167  &   4.859 &  3.366 &  0.773 &  0.399 &  2.884 \\       
11500 &   3.420 &  2.596 &  0.512 &  0.287 &  2.150  &   4.852 &  3.341 &  0.760 &  0.384 &  2.867 \\       
12000 &   3.412 &  2.575 &  0.503 &  0.279 &  2.135  &   4.845 &  3.319 &  0.750 &  0.378 &  2.853 \\       
12500 &   3.403 &  2.559 &  0.498 &  0.280 &  2.124  &   4.835 &  3.303 &  0.744 &  0.379 &  2.842 \\       
13000 &   3.389 &  2.544 &  0.493 &  0.283 &  2.114  &   4.821 &  3.288 &  0.739 &  0.383 &  2.832 \\       
13500 &   3.368 &  2.531 &  0.490 &  0.287 &  2.103  &   4.800 &  3.276 &  0.735 &  0.388 &  2.822 \\       
14000 &   3.340 &  2.520 &  0.489 &  0.292 &  2.094  &   4.771 &  3.266 &  0.733 &  0.393 &  2.813 \\       
14500 &   3.309 &  2.511 &  0.488 &  0.298 &  2.086  &   4.739 &  3.258 &  0.733 &  0.399 &  2.804 \\       
15000 &   3.277 &  2.504 &  0.488 &  0.304 &  2.078  &   4.706 &  3.252 &  0.732 &  0.405 &  2.797 \\       
15500 &   3.244 &  2.498 &  0.488 &  0.311 &  2.072  &   4.673 &  3.246 &  0.733 &  0.413 &  2.790 \\       
16000 &   3.212 &  2.492 &  0.488 &  0.318 &  2.066  &   4.641 &  3.242 &  0.733 &  0.420 &  2.784 \\       
16500 &   3.181 &  2.487 &  0.489 &  0.326 &  2.060  &   4.609 &  3.237 &  0.733 &  0.427 &  2.778 \\       
17000 &   3.151 &  2.482 &  0.489 &  0.333 &  2.055  &   4.579 &  3.233 &  0.733 &  0.434 &  2.773 \\ \hline
\end{tabular}
\end{table*}

\newpage

\begin{table*}

\caption[]{\UVEX/\IPHAS colour indices $(U-g)$, $(g-r)$, $(HeI-r)$
  $(r-H\alpha)$ and $(r-i)$ for log(g)=8.0 Koester DA white dwarfs
  including reddening}
\begin{tabular}{p{3mm}p{9mm}p{9mm}p{9mm}p{9mm}p{9mm}p{9mm}p{7mm}p{9mm}p{9mm}p{7mm}p{7mm}p{7mm}p{9mm}p{7mm}p{7mm}}
\hline 
\multicolumn{1}{l}{T (K)}&
\multicolumn{5}{l}{E(B--V)=0.0}&
\multicolumn{5}{l}{E(B--V)=1.0}&
\multicolumn{5}{l}{E(B--V)=2.0}\\
 & $U$$-$$g$ & $g$$-$$r$ & $He${\sc i}$-$$r$ & $r$$-$$H\alpha$ & $r$$-$$i$ & $U$$-$$g$ & $g$$-$$r$
 & $He${\sc i}$-$$r$ & $r$$-$$H\alpha$ & $r$$-$$i$ & $U$$-$$g$ &
$g$$-$$r$ & $He${\sc i}$-$$r$ &
 $r$$-$$H\alpha$ & $r$$-$$i$ \\\hline
 6000 & --0.141 &  0.591 &  0.102 &  0.257 &  0.378  &    1.115 &  1.482 &  0.275 &  0.429 &  1.074  &   2.448 &  2.306 &  0.483 &  0.566 &  1.775 \\
 7000 & --0.403 &  0.423 &  0.072 &  0.191 &  0.275  &    0.844 &  1.325 &  0.238 &  0.369 &  0.974  &   2.171 &  2.158 &  0.439 &  0.513 &  1.677 \\
 8000 & --0.517 &  0.312 &  0.047 &  0.101 &  0.200  &    0.726 &  1.219 &  0.208 &  0.286 &  0.901  &   2.049 &  2.056 &  0.404 &  0.435 &  1.606 \\
 9000 & --0.541 &  0.228 &  0.024 & --0.009 &  0.139  &    0.700 &  1.136 &  0.179 &  0.180 &  0.841  &   2.022 &  1.973 &  0.370 &  0.335 &  1.548 \\
10000 & --0.524 &  0.160 &  0.000 & --0.117 &  0.083  &    0.717 &  1.066 &  0.151 &  0.078 &  0.788  &   2.039 &  1.900 &  0.336 &  0.239 &  1.497 \\
11000 & --0.521 &  0.105 & --0.018 & --0.183 &  0.035  &    0.720 &  1.010 &  0.129 &  0.017 &  0.742  &   2.043 &  1.841 &  0.310 &  0.181 &  1.453 \\
12000 & --0.537 &  0.063 & --0.028 & --0.207 &  0.002  &    0.703 &  0.968 &  0.116 & --0.005 &  0.710  &   2.024 &  1.799 &  0.294 &  0.162 &  1.422 \\
13000 & --0.561 &  0.028 & --0.033 & --0.208 & --0.022  &    0.677 &0.935 &  0.109 &--0.005 &  0.687  &   1.997 &  1.768 &  0.286 &  0.164 &  1.400 \\
14000 & --0.602 & --0.004 & --0.037 & --0.205 & --0.042  &    0.634 &  0.905 &  0.104 & --0.001 &  0.667  &   1.952 &  1.741 &  0.281 &  0.168 &  1.379 \\
15000 & --0.658 & --0.030 & --0.037 & --0.196 & --0.057  &    0.575 &  0.883 &  0.104 &  0.008 &  0.651  &   1.892 &  1.722 &  0.279 &  0.178 &  1.364 \\
16000 & --0.718 & --0.051 & --0.037 & --0.184 & --0.069  &    0.514 &  0.866 &  0.104 &  0.021 &  0.639  &   1.828 &  1.707 &  0.280 &  0.191 &  1.351 \\
17000 & --0.775 & --0.069 & --0.036 & --0.170 & --0.078  &    0.454 &  0.851 &  0.104 &  0.034 &  0.629  &   1.767 &  1.695 &  0.280 &  0.204 &  1.341 \\
18000 & --0.829 & --0.084 & --0.036 & --0.158 & --0.087  &    0.398 &  0.838 &  0.105 &  0.047 &  0.621  &   1.710 &  1.685 &  0.280 &  0.217 &  1.332 \\
19000 & --0.879 & --0.098 & --0.036 & --0.147 & --0.094  &    0.347 &  0.827 &  0.105 &  0.057 &  0.613  &   1.658 &  1.675 &  0.280 &  0.227 &  1.324 \\
20000 & --0.924 & --0.110 & --0.036 & --0.138 & --0.101  &    0.301 &  0.816 &  0.105 &  0.067 &  0.606  &   1.610 &  1.666 &  0.280 &  0.237 &  1.317 \\
22000 & --1.004 & --0.132 & --0.036 & --0.124 & --0.114  &    0.219 &  0.798 &  0.104 &  0.081 &  0.593  &   1.527 &  1.651 &  0.279 &  0.251 &  1.304 \\
24000 & --1.072 & --0.151 & --0.037 & --0.114 & --0.125  &    0.149 &  0.781 &  0.103 &  0.091 &  0.581  &   1.456 &  1.637 &  0.278 &  0.262 &  1.292 \\
26000 & --1.132 & --0.170 & --0.037 & --0.106 & --0.136  &    0.087 &  0.766 &  0.103 &  0.100 &  0.571  &   1.392 &  1.623 &  0.277 &  0.270 &  1.281 \\
28000 & --1.188 & --0.185 & --0.037 & --0.095 & --0.145  &    0.030 &  0.752 &  0.102 &  0.111 &  0.561  &   1.333 &  1.612 &  0.277 &  0.282 &  1.271 \\
30000 & --1.239 & --0.198 & --0.037 & --0.078 & --0.152  &   --0.023 &  0.742 &  0.103 &  0.127 &  0.553  &   1.279 &  1.603 &  0.277 &  0.298 &  1.263 \\
35000 & --1.322 & --0.219 & --0.035 & --0.038 & --0.163  &   --0.108 &  0.725 &  0.105 &  0.167 &  0.542  &   1.192 &  1.590 &  0.279 &  0.337 &  1.252 \\
40000 & --1.362 & --0.231 & --0.034 & --0.017 & --0.169  &   --0.150 &  0.716 &  0.106 &  0.188 &  0.536  &   1.149 &  1.583 &  0.280 &  0.359 &  1.245 \\
45000 & --1.386 & --0.239 & --0.034 & --0.004 & --0.172  &   --0.175 &  0.709 &  0.106 &  0.201 &  0.532  &   1.123 &  1.578 &  0.281 &  0.372 &  1.241 \\
50000 & --1.403 & --0.244 & --0.034 &  0.005 & --0.175  &   --0.193 &  0.704 &  0.106 &  0.210 &  0.529  &   1.104 &  1.574 &  0.281 &  0.381 &  1.238 \\
55000 & --1.416 & --0.249 & --0.034 &  0.012 & --0.177  &   --0.206 &  0.700 &  0.107 &  0.217 &  0.527  &   1.090 &  1.570 &  0.281 &  0.387 &  1.236 \\
60000 & --1.427 & --0.253 & --0.033 &  0.017 & --0.179  &   --0.217 &  0.697 &  0.107 &  0.222 &  0.525  &   1.079 &  1.568 &  0.281 &  0.392 &  1.234 \\
65000 & --1.435 & --0.256 & --0.033 &  0.021 & --0.181  &   --0.226 &  0.694 &  0.107 &  0.226 &  0.523  &   1.070 &  1.565 &  0.282 &  0.397 &  1.232 \\
70000 & --1.443 & --0.259 & --0.033 &  0.025 & --0.183  &   --0.234 &  0.692 &  0.107 &  0.230 &  0.522  &   1.062 &  1.563 &  0.282 &  0.400 &  1.230 \\
75000 & --1.449 & --0.262 & --0.033 &  0.027 & --0.184  &   --0.241 &  0.690 &  0.107 &  0.232 &  0.520  &   1.055 &  1.562 &  0.281 &  0.403 &  1.229 \\
80000 & --1.455 & --0.264 & --0.033 &  0.030 & --0.185  &   --0.247 &  0.688 &  0.107 &  0.235 &  0.519  &   1.049 &  1.560 &  0.281 &  0.405 &  1.227 \\ \hline    
\end{tabular}
\end{table*}

\newpage

\addtocounter{table}{-1}
\begin{table*}
\caption[]{, continued}
\flushleft
\begin{tabular}{cccccccccccccccc}
\hline 
\multicolumn{1}{l}{T (K)}&
\multicolumn{5}{l}{E(B--V)=3.0}&
\multicolumn{5}{l}{E(B--V)=4.0}\\
      & $U$$-$$g$ & $g$$-$$r$ & $He${\sc i}$-$$r$ & $r$$-$$H\alpha$ &
$r$$-$$i$ &$U$$-$$g$ & $g$$-$$r$ & $He${\sc i}$-$$r$ & $r$$-$$H\alpha$ & $r$$-$$i$\\ \hline
 6000 &  3.840 &  3.084 &  0.724 &  0.671 &  2.480  &   5.272 &  3.834 &  0.996 &  0.744 &  3.189 \\
 7000 &  3.559 &  2.942 &  0.674 &  0.624 &  2.384  &   4.990 &  3.695 &  0.941 &  0.703 &  3.094 \\
 8000 &  3.435 &  2.841 &  0.634 &  0.551 &  2.314  &   4.866 &  3.594 &  0.896 &  0.634 &  3.026 \\
 9000 &  3.407 &  2.757 &  0.595 &  0.456 &  2.259  &   4.838 &  3.508 &  0.853 &  0.544 &  2.972 \\
10000 &   3.426 &  2.681 &  0.556 &  0.364 &  2.210  &   4.857 &  3.429 &  0.809 &  0.457 &  2.925\\
11000 &   3.430 &  2.620 &  0.526 &  0.311 &  2.167  &   4.862 &  3.366 &  0.775 &  0.407 &  2.884\\
12000 &   3.411 &  2.577 &  0.507 &  0.294 &  2.137  &   4.844 &  3.322 &  0.755 &  0.393 &  2.855\\
13000 &   3.383 &  2.547 &  0.498 &  0.297 &  2.115  &   4.815 &  3.292 &  0.744 &  0.397 &  2.834\\
14000 &   3.337 &  2.521 &  0.492 &  0.303 &  2.095  &   4.768 &  3.268 &  0.737 &  0.404 &  2.814\\
15000 &   3.274 &  2.505 &  0.490 &  0.313 &  2.079  &   4.705 &  3.253 &  0.735 &  0.414 &  2.798\\
16000 &   3.209 &  2.493 &  0.490 &  0.326 &  2.067  &   4.639 &  3.243 &  0.735 &  0.427 &  2.785\\
17000 &   3.147 &  2.483 &  0.490 &  0.339 &  2.056  &   4.576 &  3.234 &  0.735 &  0.441 &  2.774\\
18000 &   3.089 &  2.474 &  0.491 &  0.352 &  2.047  &   4.517 &  3.227 &  0.735 &  0.453 &  2.765\\
19000 &   3.036 &  2.466 &  0.491 &  0.363 &  2.039  &   4.463 &  3.220 &  0.735 &  0.464 &  2.756\\
20000 &   2.988 &  2.459 &  0.490 &  0.372 &  2.031  &   4.415 &  3.213 &  0.734 &  0.474 &  2.749\\
22000 &   2.903 &  2.445 &  0.489 &  0.386 &  2.018  &   4.329 &  3.202 &  0.733 &  0.488 &  2.735\\
24000 &   2.831 &  2.433 &  0.488 &  0.397 &  2.006  &   4.256 &  3.191 &  0.731 &  0.500 &  2.723\\
26000 &   2.766 &  2.421 &  0.487 &  0.407 &  1.995  &   4.192 &  3.180 &  0.730 &  0.509 &  2.712\\
28000 &   2.707 &  2.411 &  0.486 &  0.418 &  1.985  &   4.131 &  3.171 &  0.729 &  0.521 &  2.702\\
30000 &   2.652 &  2.404 &  0.486 &  0.434 &  1.977  &   4.076 &  3.166 &  0.729 &  0.537 &  2.693\\
35000 &   2.562 &  2.395 &  0.489 &  0.473 &  1.964  &   3.985 &  3.158 &  0.732 &  0.576 &  2.681\\
40000 &   2.518 &  2.389 &  0.490 &  0.495 &  1.958  &   3.940 &  3.154 &  0.733 &  0.597 &  2.674\\
45000 &   2.492 &  2.384 &  0.490 &  0.508 &  1.953  &   3.913 &  3.150 &  0.733 &  0.610 &  2.669\\
50000 &   2.473 &  2.381 &  0.491 &  0.516 &  1.950  &   3.894 &  3.148 &  0.734 &  0.619 &  2.666\\
55000 &   2.458 &  2.379 &  0.491 &  0.523 &  1.948  &   3.879 &  3.145 &  0.734 &  0.625 &  2.663\\
60000 &   2.447 &  2.376 &  0.491 &  0.528 &  1.946  &   3.867 &  3.144 &  0.734 &  0.631 &  2.661\\
65000 &   2.437 &  2.375 &  0.491 &  0.532 &  1.944  &   3.858 &  3.142 &  0.734 &  0.635 &  2.660\\
70000 &   2.429 &  2.373 &  0.491 &  0.536 &  1.942  &   3.849 &  3.141 &  0.734 &  0.638 &  2.658\\
75000 &   2.422 &  2.371 &  0.491 &  0.539 &  1.941  &   3.842 &  3.139 &  0.734 &  0.641 &  2.656\\
80000 &   2.416 &  2.370 &  0.491 &  0.541 &  1.940  &   3.836 &
3.138 &  0.734 &  0.643 &  2.655\\ \hline
\end{tabular}
\end{table*}

\clearpage
\newpage

\begin{table*}
\caption[]{\UVEX/\IPHAS colour indices $(U-g)$, $(g-r)$, $(HeI-r)$ $(r-H\alpha)$ and $(r-i)$ for log(g)=8.0 Koester DB white dwarfs including reddening.}
\begin{tabular}{p{3mm}p{9mm}p{9mm}p{9mm}p{7mm}p{9mm}p{9mm}p{7mm}p{9mm}p{7mm}p{7mm}p{7mm}p{7mm}p{9mm}p{7mm}p{7mm}}
\hline 
\multicolumn{1}{l}{T (K)}&
\multicolumn{5}{l}{E(B--V)=0.0}&
\multicolumn{5}{l}{E(B--V)=1.0}&
\multicolumn{5}{l}{E(B--V)=2.0}\\
      & $U$$-$$g$ & $g$$-$$r$ & $He${\sc i}$-$$r$ & $r$$-$$H\alpha$ & $r$$-$$i$ & $U$$-$$g$ & $g$$-$$r$ &
$He${\sc i}$-$$r$ & $r$$-$$H\alpha$ & $r$$-$$i$ & $U$$-$$g$ &
$g$$-$$r$ & $He${\sc i}$-$$r$ & $r$$-$$H\alpha$
& $r$$-$$i$ \\ \hline
10000 & --0.787 &  0.162 &  0.033 &  0.172 &  0.081  &    0.444 &  1.083 &  0.189 &  0.361 &  0.781  &   1.757 &  1.930 &  0.380 &  0.515 &  1.485  \\
11000 & --0.874 &  0.109 &  0.031 &  0.161 &  0.049  &    0.353 &  1.033 &  0.186 &  0.351 &  0.749  &   1.664 &  1.883 &  0.375 &  0.507 &  1.453  \\
12000 & --0.944 &  0.067 &  0.037 &  0.153 &  0.023  &    0.281 &  0.995 &  0.190 &  0.345 &  0.724  &   1.590 &  1.848 &  0.378 &  0.502 &  1.428  \\
13000 & --0.993 &  0.032 &  0.050 &  0.146 & --0.000  &    0.231 &  0.962 &  0.202 &  0.339 &  0.701  &   1.539 &  1.816 &  0.388 &  0.498 &  1.406  \\
14000 & --1.029 &  0.001 &  0.072 &  0.140 & --0.020  &    0.193 &  0.933 &  0.223 &  0.334 &  0.681  &   1.500 &  1.788 &  0.408 &  0.494 &  1.386  \\
15000 & --1.053 & --0.024 &  0.101 &  0.135 & --0.037  &    0.168 &  0.908 &  0.250 &  0.330 &  0.665  &   1.475 &  1.764 &  0.435 &  0.491 &  1.370  \\
16000 & --1.068 & --0.045 &  0.134 &  0.130 & --0.050  &    0.153 &  0.888 &  0.283 &  0.326 &  0.652  &   1.460 &  1.744 &  0.466 &  0.488 &  1.358  \\
17000 & --1.077 & --0.063 &  0.167 &  0.125 & --0.060  &    0.144 &  0.870 &  0.314 &  0.322 &  0.643  &   1.451 &  1.725 &  0.497 &  0.485 &  1.349  \\
18000 & --1.083 & --0.078 &  0.194 &  0.120 & --0.068  &    0.138 &  0.855 &  0.341 &  0.318 &  0.635  &   1.445 &  1.709 &  0.522 &  0.481 &  1.343  \\
19000 & --1.089 & --0.091 &  0.212 &  0.115 & --0.075  &    0.132 &  0.841 &  0.358 &  0.314 &  0.629  &   1.440 &  1.695 &  0.539 &  0.478 &  1.337  \\
20000 & --1.095 & --0.103 &  0.220 &  0.111 & --0.081  &    0.126 &  0.830 &  0.366 &  0.310 &  0.624  &   1.433 &  1.684 &  0.546 &  0.475 &  1.332  \\
22000 & --1.105 & --0.117 &  0.216 &  0.107 & --0.088  &    0.115 &  0.816 &  0.361 &  0.306 &  0.617  &   1.423 &  1.670 &  0.541 &  0.471 &  1.325  \\
24000 & --1.118 & --0.128 &  0.199 &  0.104 & --0.094  &    0.102 &  0.807 &  0.343 &  0.304 &  0.611  &   1.408 &  1.662 &  0.523 &  0.469 &  1.319  \\
26000 & --1.141 & --0.142 &  0.185 &  0.100 & --0.103  &    0.077 &  0.793 &  0.329 &  0.301 &  0.602  &   1.383 &  1.650 &  0.508 &  0.467 &  1.310  \\
28000 & --1.169 & --0.156 &  0.175 &  0.098 & --0.113  &    0.048 &  0.781 &  0.319 &  0.299 &  0.592  &   1.353 &  1.639 &  0.497 &  0.465 &  1.300  \\
30000 & --1.199 & --0.169 &  0.164 &  0.096 & --0.122  &    0.018 &  0.770 &  0.307 &  0.297 &  0.583  &   1.321 &  1.630 &  0.485 &  0.464 &  1.291  \\
35000 & --1.270 & --0.196 &  0.134 &  0.092 & --0.139  &   --0.056 &  0.747 &  0.277 &  0.294 &  0.565  &   1.244 &  1.611 &  0.454 &  0.461 &  1.274  \\
40000 & --1.329 & --0.218 &  0.116 &  0.088 & --0.154  &   --0.117 &  0.728 &  0.258 &  0.290 &  0.551  &   1.182 &  1.594 &  0.435 &  0.459 &  1.259  \\
50000 & --1.463 & --0.236 &  0.038 &  0.080 & --0.180  &   --0.256 &  0.717 &  0.180 &  0.284 &  0.524  &   1.039 &  1.589 &  0.355 &  0.453 &  1.232  \\ \hline    
\end{tabular} 
\end{table*}

\newpage
\addtocounter{table}{-1}
\begin{table*}
\caption[]{, continued}
\flushleft
\begin{tabular}{ccccccccccc} \hline
\multicolumn{1}{l}{T (K)}&
\multicolumn{5}{l}{E(B--V)=3.0}&
\multicolumn{5}{l}{E(B--V)=4.0}\\
 &$U$$-$$g$ & $g$$-$$r$ & $He${\sc i}$-$$r$ & $r$$-$$H\alpha$ &
$r$$-$$i$ & $U$$-$$g$ & $g$$-$$r$ & $He${\sc i}$-$$r$ & $r$$-$$H\alpha$ & $r$$-$$i$\\ \hline
10000 &  3.136 &  2.722 &  0.605 &  0.635 &  2.193  &   4.561 &  3.481 &  0.863 &  0.723 &  2.904\\
11000 &  3.041 &  2.678 &  0.598 &  0.629 &  2.162  &   4.466 &  3.438 &  0.854 &  0.719 &  2.874\\
12000 &  2.967 &  2.645 &  0.600 &  0.626 &  2.137  &   4.391 &  3.406 &  0.854 &  0.716 &  2.849\\
13000 &  2.915 &  2.614 &  0.609 &  0.622 &  2.114  &   4.339 &  3.375 &  0.863 &  0.714 &  2.827\\
14000 &  2.875 &  2.587 &  0.627 &  0.620 &  2.095  &   4.299 &  3.348 &  0.880 &  0.713 &  2.808\\
15000 &  2.850 &  2.563 &  0.653 &  0.618 &  2.080  &   4.273 &  3.324 &  0.904 &  0.712 &  2.793\\
16000 &  2.835 &  2.542 &  0.684 &  0.616 &  2.068  &   4.259 &  3.302 &  0.934 &  0.711 &  2.781\\
17000 &  2.827 &  2.523 &  0.713 &  0.613 &  2.060  &   4.252 &  3.282 &  0.963 &  0.709 &  2.774\\
18000 &  2.822 &  2.505 &  0.738 &  0.610 &  2.054  &   4.248 &  3.264 &  0.988 &  0.707 &  2.768\\
19000 &  2.817 &  2.491 &  0.754 &  0.607 &  2.048  &   4.244 &  3.248 &  1.003 &  0.704 &  2.763\\
20000 &  2.811 &  2.479 &  0.761 &  0.605 &  2.044  &   4.237 &  3.236 &  1.009 &  0.702 &  2.759\\
22000 &  2.800 &  2.465 &  0.755 &  0.602 &  2.037  &   4.226 &  3.222 &  1.002 &  0.699 &  2.752\\
24000 &  2.785 &  2.458 &  0.737 &  0.600 &  2.031  &   4.211 &  3.215 &  0.984 &  0.698 &  2.747\\
26000 &  2.759 &  2.447 &  0.722 &  0.598 &  2.022  &   4.185 &  3.205 &  0.969 &  0.696 &  2.738\\
28000 &  2.728 &  2.437 &  0.710 &  0.597 &  2.012  &   4.153 &  3.196 &  0.957 &  0.696 &  2.728\\
30000 &  2.695 &  2.429 &  0.698 &  0.596 &  2.003  &   4.119 &  3.189 &  0.944 &  0.695 &  2.719\\
35000 &  2.616 &  2.413 &  0.666 &  0.594 &  1.986  &   4.039 &  3.176 &  0.911 &  0.694 &  2.701\\
40000 &  2.552 &  2.398 &  0.646 &  0.592 &  1.971  &   3.975 &  3.162 &  0.890 &  0.693 &  2.686\\
50000 &  2.405 &  2.399 &  0.566 &  0.588 &  1.943  &   3.824 &  3.168
&  0.810 &  0.689 &  2.658\\ \hline
\end{tabular}
\end{table*}

\end{document}